\documentclass[]{aa}  
\usepackage[colorlinks=true,linkcolor=blue,citecolor=blue]{hyperref} %changed for arxiv
\usepackage{graphicx}

\usepackage{txfonts}
\usepackage{subfig}
\usepackage{xcolor,colortbl}

\usepackage[]{natbib}
\usepackage{multirow}
\usepackage{amsmath}
\usepackage{enumitem}

\usepackage[pdf]{pstricks}
\usepackage{pst-all}
\usepackage{subfig}
\usepackage{float}

\usepackage{scalerel}
\usepackage{tikz}
\usetikzlibrary{shapes.geometric, arrows}

\def\deg{$^{\circ}$}
\def\degb{^{\circ}}

\begin{document}

   \title{Spatially resolved spectroscopy of the debris disk HD\,32297\thanks{Based on data collected at the European Southern Observatory, Chile under the programs 098.C-0686(A) and 098.C-0686(B)}\fnmsep\thanks{Reduced data (FITS file) of the S/N maps are only available on CDS via anonymous ftp cdsarc.u-strasbg.fr(130.79.128.5) or via {http://cdsweb.u-strasbg.fr/cgi-bin/qcat?J/A+A/}}}
   \subtitle{Further evidence of small dust grains}

   \author{T. Bhowmik\inst{1}, A. Boccaletti\inst{1}, P. Th{\'e}bault\inst{1}, Q. Kral\inst{1}, J.Mazoyer\inst{2}\fnmsep\thanks{Sagan NHFP Fellow}, J. Milli\inst{3,8}, A.L. Maire\inst{4,7}, R. G. van Holstein\inst{10}, J.-C. Augereau\inst{3}, P. Baudoz\inst{1}, M. Feldt\inst{4}, R. Galicher \inst{1}, T. Henning\inst{4}, A.-M. Lagrange\inst{3}, J. Olofsson\inst{4,6,7}, E. Pantin\inst{9}, C. Perrot\inst{1,5,6}}

   \institute{LESIA, Observatoire de Paris, Universit{\'e} PSL, CNRS, Sorbonne Universit{\'e}, Universit{\'e} de Paris, 5 place Jules Janssen, 92195 Meudon, France\\
              \email{trisha.bhowmik@obspm.fr}
         \and
                      Jet Propulsion Laboratory, California Institute of Technology, 4800 Oak Grove Drive, Pasadena, CA 91109, USA
         \and
             CNRS, IPAG, Universit{\'e} Grenoble Alpes, 38000 Grenoble, France 
        \and
             Max-Planck-Institut f{\"u}r Astronomie, K\"onigstuhl 17, 69117 Heidelberg, Germany 
        \and
            Instituto de F{\'i}sica y Astronom{\'i}a, Facultad de Ciencias, Universidad de Valpara{\'i}so, Avenue Gran Breta\~na 1111, Valpara{\'i}so, Chile 
        \and
            N{\'u}cleo Milenio Formaci{\'o}n Planetaria â NPF, Universidad de Valpara{\'i}so, Avenue Gran Breta\~na 1111, Valpara{\'i}so, Chile 
        \and
            STAR Institute, Universit\'e de Li\`ege, All\'ee du Six Ao\^ut 19c, B-4000 Li\`ege, Belgium
        \and
            European Southern Observatory, Alonso de C\'ordova 3107, Casilla 19001 Vitacura, Santiago 19, Chile 
        \and
            Laboratoire AIM, CEA/DRF - CNRS - Universit\'e Paris Diderot, IRFU/SAp, UMR 7158, 91191 Gif sur Yvette, France 
        \and
            Leiden Observatory, Leiden University, PO Box 9513, 2300 RA Leiden, The Netherlands 
             }

   \date{}
\titlerunning{HD\,32297}
\authorrunning{Bhowmik et al.}
% \abstract{}{}{}{}{} 
% 5 {} token are mandatory
 
  \abstract
   {Spectro-photometry of debris disks in total intensity and polarimetry can provide new insight into the properties of the dust grains therein (size distribution and optical properties).
   }
   {We aim to constrain the morphology of the highly inclined debris disk HD\,32297. We also intend to obtain spectroscopic and polarimetric measurements to retrieve information on the particle size distribution within the disk for certain grain compositions.}
   { We observed HD\,32297 with  SPHERE in Y, J, and H bands in total intensity and in J band in polarimetry. The observations are compared to synthetic models of debris disks and we developed methods to extract the photometry in total intensity {overcoming the} data-reduction artifacts, namely the self-subtraction. The spectro-photometric measurements averaged along the disk mid-plane are then compared to model spectra of various grain compositions. 
    
   }
   {
    These new images reveal the very inner part of the system as close as 0.15$''$.  The disk image is mostly dominated by the forward scattering making one side (half-ellipse) of the disk more visible, but observations in total intensity are deep enough to also detect the back side for the very first time. 
   The images as well as the surface brightness profiles of the disk rule out the presence of a gap as previously proposed. We do not detect any significant asymmetry between the northeast and southwest sides of the disk.
   The spectral reflectance features a {"gray to blue" color} which is interpreted as the presence of grains far below the blowout size.
   }
   {The presence of sub-micron grains in the disk is suspected to be the result of gas drag and/or "avalanche mechanisms". 
   The blue color of the disk could be further investigated with additional total intensity and polarimetric observations in K and H bands respectively to confirm the spectral slope and the fraction of polarization.
   }


   \maketitle
%
%________________________________________________________________

\section{Introduction}

Debris disks are a class of circumstellar disks in which the dust content is thought to be continuously replenished by collisions of planetesimals, usually distributed in a single belt or in multiple belts.
The dust present in these disks can be observed by scattered light imaging and/or thermal emission. Morphological and photo-spectroscopic analysis provide constraints on the spatial distribution and physical properties of grains present in a planetary system.
Even though debris disks are believed to be depleted in gas, a number of them are significantly gas rich, very likely because of second-generation production of gas released when planetesimals collide and produce the observed dust \citep{Kral2017}. 
In addition, debris disks frequently show identifiable structures like blobs
\citep[AU\,Mic,][]{Boccaletti2018}, warps \citep[$\beta$ Pic,][]{Mouillet1997}, multiple belts \citep[HIP67497 and NZ Lup]{Bonnefoy2017, Boccaletti2019}, asymmetries \citep[HD\,15115]{Mazoyer2014}, and other irregularities in the disk. These structures might be sculpted by transient breakups of massive bodies \citep{Jackson2014,Kral2015}, stellar flybys \citep{Lestrade2011}, stellar companions \citep{Thebault2012}, or interactions with the inter-stellar medium (ISM) \citep{Hines2007, Kalas2005}, or be owing to the presence of gas \citep{LyraKuchner2013} within the disk. One of the most frequently invoked explanations for the presence of most of these structures is, however, the perturbing effect of a planet \citep{Thebault2012b}. In this scenario, structures can be the telltale signature of {undetected} planets, such as in the case of $\beta$ Pictoris b, which was first inferred indirectly from a warp in the disk and later imaged directly confirming the prediction \citep{Mouillet1997, Lagrange2009}.

HD\,32297 is an A-type star \citep{Silverstone2000} ($V=8.14\pm 0.01,  J=7.687\pm0.024, H=7.624\pm0.051, K=7.594\pm0.018 $) located at 133\,pc  \citep[Gaia DR2,][]{Gaia2018}. The age of the star is estimated to be $\gtrsim$15\,Myr \citep{Rodigas2014} and $<$30\,Myr \citep{Kalas2005}.
A fractional luminosity of L$_{IR}$/L{$_\star$} $\geq$ 2.7$\times10^{-3}$ found with the Infrared Astronomical Satellite \citep[IRAS,][]{Silverstone2000} provides evidence for the presence of cold dust arranged in a belt. This high fractional luminosity makes it one of the brightest debris disks known to date \citep[e.g.,][]{Thebault2019}.
The disk was first resolved in scattered light by the Hubble Space Telescope (HST) NICMOS as an edge-on system and detected up to radial distances of 400\,au \citep{Schneider2005}. A  significant surface brightness asymmetry (southwest ansae brighter than northeast ansae) was reported and attributed tentatively to the  presence of an unseen planet. 
\cite{Kalas2005} detected the disk at larger separations (400 to 1680\,au) from the ground with Keck in the R band and measured a blue color when comparing to the HST near-infrared (NIR) data.  

{In apparent contradiction to the scattered light observations of \cite{Kalas2005} and \cite{Schneider2005}, mid-infrared (MIR) observations published by \cite{Moerchen2007} and \cite{Fitzgerald2007} show that the northeast side appears brighter than the southwest side beyond $0.75''$, although the impact of the angular resolution and signal to noise ratio (S/N) in these images could be questionable.} 
\cite{Maness2008} and \cite{Mawet2009} observed a similar brightness asymmetry to \cite{Schneider2005} and \cite{Kalas2005} in millimeter emission and K band respectively, but still with low angular resolution. 

At large distances, as observed with Keck and confirmed with HST/NICMOS, the disk appears bowed, which has been interpreted as the possible interaction of small dust grains residing at large distances with the ISM \citep{Debes2009}. \cite{Rodigas2014} confirmed the bow shape of the disk at the L$'$ band. Recently, \cite{LeeChiang2016} showed that similar structures could also be the result of an interaction between a planet in an eccentric orbit and planetesimals perturbed by radiation pressure. The presence of {small dust} grains in a large halo has been confirmed with very deep HST/STIS observations \citep{Schneider2014}.

With the advance of high-contrast imaging, \cite{Boccaletti2012}, \cite{Currie2012} and \cite{Esposito2014} were able to resolve the disk in the NIR as an inclined belt ($\sim$88\deg) located at $\sim$130\,au (after correcting for the new Gaia distance). 
According to \cite{Boccaletti2012}, there is no significant brightness asymmetry between the two sides of the disk in H and K bands as observed with NACO,
while \cite{Currie2012} suggest the  southwest side to be brighter than the northeast side at $r= 35-80$\,au.

High-contrast polarimetry combined with total intensity in the NIR was first achieved by \cite{Asensio-Torres2016} with Subaru/HiCIAO observations, and with the aim to break degeneracies on the geometrical parameters and grain  properties. They reported a gap in the total intensity H band data, which was not visible in their polarimetric data.

Finally, the most recent observations were carried out with ALMA by \citet[][hereafter MG18]{MacGregor2018} who concluded that at millimeter wavelengths the disk is composed of a planetesimal belt with an inner edge at 78 au and a outer edge at 122\,au, and an extended halo up to 440\,au. The presence of millimeter grains in the halo complicates the understanding of the large bow, which is expected to be populated with loosely bound grains placed on very eccentric orbits by stellar radiation pressure \citep{StrubbeChiang06,ThebaultWu08} that should be sensitive to interactions with the ISM. We note however that there are some other alternatives and the ALMA halo may be explained as the presence of a scattered disk \citep{Geiler2019}. 

Furthermore, CO gas emission \citep[][MG18]{Greaves2016} was detected with ALMA corresponding to a total mass of $\sim 7\times10^{-2}$\,$M_{\oplus}$ derived from an optically thin CO isotopolog  \citep[][submitted]{Moor2019}. This large quantity of gas could interact and drag the dust in this system \citep{Takeuchi2001}. The smallest bound grains would be pushed backwards and the unbound grains would be slowed down and may accumulate in greater quantity than in gas-depleted systems.

Modeling of the spectral energy distribution (SED) suggests that the main belt is populated with sub-micron grains \citep{Fitzgerald2007, Currie2012}, which is in agreement with the color index measured between visible and NIR by \cite{Kalas2005}. The presence of such sub-micron grains is unexpected, because they should be smaller than the blowout size due to radiation pressure, and should therefore be ejected on very short timescales \citep{Kral2013}. Moreover, taking into account the size of the belt inferred by \citet{Boccaletti2012}
and using photometry from Herschel, \cite{Donaldson2013} were able to draw constraints on the grain composition. These latter authors concluded that the system is made of an inner warm belt at $\geq$1.1 au, and an outer ring at 110\,au populated with $\geq2.2\muup$m, high-porosity grains consistent with cometary-like composition. However, this result was questioned by \cite{Rodigas2014} who favored pure icy grains instead.

 In this paper we present new high-contrast imaging observations of HD\,32297 with Spectro-Polarimetric High-contrast Exoplanet Research \citep[SPHERE,][]{Beuzit2019} at VLT in Chile, in the NIR from Y to H band complemented with polarimetric observations in the J band. With the new observations at NIR wavelengths, we resolve the disk at separations as close as 0.15$''$ and perform a spectral analysis.
We briefly discuss the observational technique and the data reduction for total intensity and polarimetric data in Sect. \ref{sec:obs}. The disk features seen in the images are described in Sect. \ref{sec:morpho}. 
The procedure of the modeling and parametric study is presented in Sect. \ref{sec:model}. 
In Sect. \ref{sec:photom}, we describe how the photometry is retrieved from the images in the various spectral channels, for both total intensity and polarimetry.
The resulting spectrum is compared to various grain models in Sect. \ref{sec:grains}. 
Finally, we discuss the implication of the colors in terms of grain sizes in Sect. \ref{sec:discuss}.

%%%%%%%%%%%%%%%%%%%%%%%%%%%%%%%%%%%%%%%%%%%
%%%%%%%%%%%%%%%%%%%%%%%%%%%%%%%%%%%%%%%%%%%
\section{Observation and data reduction}
\label{sec:obs}

\begin{table*}   
\centering   
\caption{SPHERE observation log} 
\begin{tabular}{c c c c c c c c c c}     % 7 columns 
\hline\hline                             % To combine 4 columns into a single one 
Data UT & prog. ID & Filter & PC & Field rotation ($\degb$) & DIT (s)  & T$_{\mathrm{exp}}$ (s) & Seeing ($''$) & $\tau_0$ (ms) & TN ($\degb$)  \\ 
\hline                    
   2016-12-19 & 098.C-0686(A) & IRDIS-BB\_H & &25.16  & 64 & 7168 &0.72 & 8.8 & -1.75 \\  
   2016-12-19 & 098.C-0686(A) & IFS-YJ & &25.42 & 64  & 7168 &0.72 & 8.7 & -1.75 \\
   2016-12-16 & 098.C-0686(B) & IRDIS-DPI-BB\_J & 10& stabilized & 64 & 5120 & 0.47& 8.0& -1.7\\
\hline   
\end{tabular}
\tablefoot{From left to right: the observation data, program ID, filter combination, the number of polarimetric cycle (PC), the total field rotation in degrees, the individual integration time of each frame (DIT) in seconds, the true time in seconds ($T_{exp}$), the DIMM seeing in arcseconds, $\tau_0$ the correlation time in milleseconds and the true north correction angle in degrees (TN)}
\label{table:obs_log}
\end{table*}

\subsection{SPHERE} % change this subtitle !
%%% general : no need for bold in acronyms

{SPHERE is an instrument} installed at the VLT for high-contrast direct imaging of giant planets around young nearby stars \citep{Beuzit2019}.
Its combination of extreme adaptive optics \citep[AO,][]{Fusco2014} and advanced coronagraphy has allowed
observation of {circumstellar} disks from the ground. SPHERE consists of three instruments: the Infra-Red Dual-beam Imager and Spectrograph \citep[IRDIS,][]{Dohlen2008}, the Integral Field Spectrograph \citep[IFS,][]{Claudi2008}, and the Zurich  IMaging  POLarimeter \citep[ZIMPOL,][]{Thalmann2008}. IRDIS is a dual band imager using two narrowband or broadband filters in the Y, J, H, or K bands \citep{Vigan2010}. The IFS is a spectro-imager which delivers 39 simultaneous images across the YJ (IRDIFS mode) or YJH (IRDIFS-ext mode) bands. These two instruments can be used for parallel observations with the IRDIFS mode. 

\subsection{Observations}
HD\,32297 was observed with SPHERE on  December 19$^{}$, 2016, in the IRDIFS mode using pupil stabilization in order to {take advantage of angular differential imaging \citep[ADI,][]{Marois2006ADI} for calibration of stellar residuals during post processing}. An apodized Lyot coronagraph N\_ALC\_YJH\_S \citep{Carbillet2011} with a diameter of 185 mas allows attenuation of the starlight in the AO corrected radius (0.84$''$ at 1.65$\muup$m).

For IRDIS the observations were performed in the broadband H filter (1.625 $\muup$m central wavelength, 0.29 $\muup$m filter width). IRDIS has a field of view (FOV) of $11''\times11''$ and pixel size of 12.25$\pm$0.02 mas. For IFS we used the YJ mode (0.95-1.35 $\muup$m), which provides a spectral resolution R$\simeq$54. The IFS has a FOV of $1.73'' \times 1.73''$ and a pixel size of 7.46$\pm$0.02 mas \citep{Maire2016}. The atmospheric conditions were good with seeing below 0.8$''$ and correlation time of about $8-9$\,ms. The observing sequence is as follows: point-spread function (PSF; with the star outside the {coronagraphic mask with} a neutral density ND2), star center \citep[coronagraphic image with four crosswise replicas of the star created with the deformable mirror to monitor,][]{Langlois2013}, long science coronagraphic exposures, second PSF and sky background. Further details on the observation log are provided in Table \ref{table:obs_log}.

Furthermore, HD\,32297 was observed with IRDIS dual polarimetric imaging (DPI) mode a few days apart in field stabilized mode to measure the polarized flux of the  disk. Several polarization cycles were taken in the J band (1.25 $\muup$m), each consisting of exposures at four orientations of the half wave plate ($0\degb, 22.5\degb, 45\degb, 67.5\degb$) to measure the full Stokes parameters. The observing sequence is similar to that of the IRDIFS mode for IRDIS; the two filters of IRDIS are replaced with polarizers to split the polarization into two orthogonal directions \citep{Langlois2014}.

\subsection{IRDIFS data reduction}

The preliminary data reduction, including flat-field corrections, sky and dark subtractions, star-centering using waffle pattern,
bad-pixel removal, distortion correction \citep{Maire2016}, and wavelength calibration, is done at the SPHERE Data Centre\footnote{http://sphere.osug.fr} \citep{Delorme2017} using the data reduction and handling (DRH) pipeline \citep{Pavlov2008,Mesa2015} for the total intensity data obtained in IRDIFS mode. This step provides a data cube which is further processed with ADI techniques based on several algorithms, such as Karhunen-Lo\`eve Image Projection \citep[KLIP,][]{Soummer2012}, classical angular differential imaging \citep[cADI,][]{Marois2006ADI}, and template locally optimized combination of images \citep[TLOCI,][]{Marois2015}, using two pipelines, namely the SpeCal pipeline \citep{Galicher2018} and another one developed by \cite{Boccaletti2015}. 

Karhunen-Lo\`eve Image Projection is an algorithm based on the principle component analysis (PCA). The algorithm involves reformatting the science data into a covariance matrix to remove redundancy. Eigenvalues and eigenvectors are calculated for the covariance matrix which is used to form the KL (Karhunen-Lo\`eve) basis, {which itself is truncated to a given number of modes} \citep{Soummer2012}. The reference image is built by projecting the science data onto the {truncated KL basis} and then subtracting them  out frame by frame. The result is derotated according to the parallactic angle variation as in any other ADI techniques allowing reduction of the stellar halo. The final image is normalized to the maximum of the measured PSF.

In this paper, we work with the KLIP reduced data truncated to ten modes using the pipeline developed by \cite{Boccaletti2015} as it provides the optimal S/N.

\subsection{Dual polarimetric imaging data reduction}

The stokes vectors Q and U are calculated from the polarization cycle observed through each half-wave plate position. To mitigate the instrumental polarization we use the double subtraction method \citep{Tinbergen1996}. The final Q and U vectors can be written as follows:

\begin{equation} 
\label{eq1}
\begin{aligned}
Q &= \frac{Q^+ - Q^-}{2}\\
U &= \frac{U^+ - U^-}{2}
\end{aligned}
,\end{equation}

\noindent where $Q^+, U^+, Q^-, U^{-}$ are {obtained from} observations through the half wave plate positions at $0\degb, 22.5\degb, 45\degb, 67.5\degb$.

We retrieve the $Q_\phi$ and $U_\phi$ azimuthal Stokes vectors which are expressed in polar coordinates as explained in \cite{Schmid2006}. The $Q_\phi$ and the $U_\phi$ vectors can be written as:

\begin{equation} 
\label{eq2}
\begin{aligned}
Q_\phi &= -Q \cos 2 \phi{+ U} \sin 2 \phi\\
U_\phi &= -Q \sin 2 \phi- U \cos 2 \phi
\end{aligned}
,\end{equation}

\noindent where $\phi$ is the azimuthal angle with respect to the center of the  star.  
%%%
We recover the disk signal in the $Q_\phi$ vector and the noise in the $U_\phi$ vector under the  assumption that the disk is {optically thin and undergoes} only single scattering. Further correction to instrumental offset is not done as the $U_{\phi}$ image is dominated by noise and has very low true disk signal.

%%%%%%%%%%%%%%%%%%%%%%%%%%%%%%%%%%%%%%%%%%%
%%%%%%%%%%%%%%%%%%%%%%%%%%%%%%%%%%%%%%%%%%%

\section{Morphology of the disk}
\label{sec:morpho}
\subsection{General description}
\label{sec:gen_descp}
 
\begin{figure*}
\centering
    \includegraphics[width=0.95\textwidth]{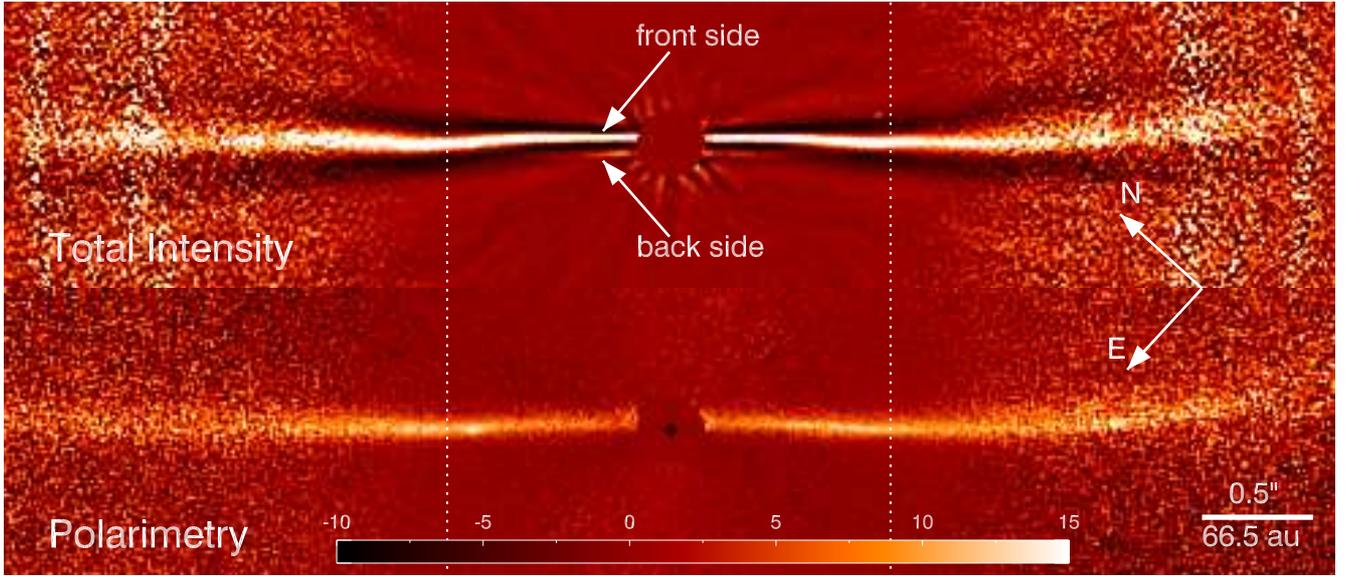}
\caption{\textit{Top:} S/N map of total intensity with IRDIS in BB\_H filter. \textit{Bottom:} S/N map of polarimetric image observed with IRDIS in BB\_J filter. The dashed line indicates approximate position of the ansa in both images and the arrows indicate front and  back sides of the disk. Both images are rotated to 90\deg\,-\,PA and cropped at {$6''\times1.3''$}. The color bar shows the intensity in both the images.}
    \label{fig:SNR}
\end{figure*}

\begin{figure}
        \includegraphics[width=\hsize]{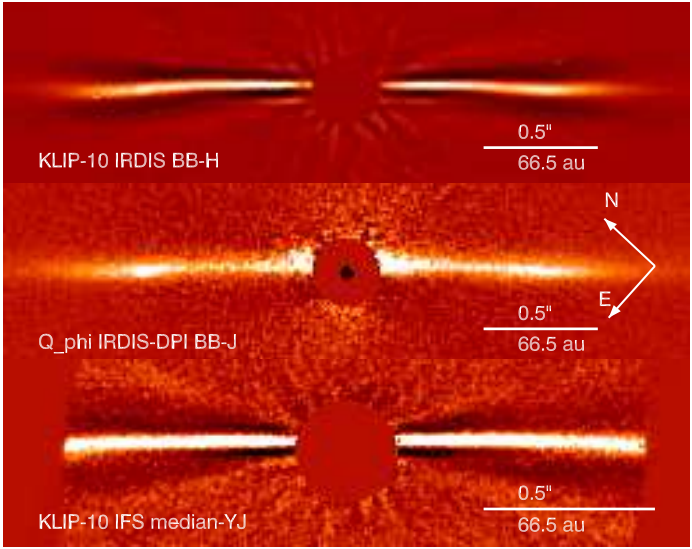}
\caption{ Inner part of the disk with IRDIS in BB\_H (KLIP, top), IRDIS in DPI ($Q_{\phi}$, middle), and IFS (YJ combined, bottom). The top two images are cropped at 3$''\times$0.8$''$ and the IFS image is cropped at 2$''\times$0.5$''$. All images are rotated to 90\deg\,-\,PA. All images are scaled linearly in the range [$-1\times10^{-5}$,$1\times10^{-5}$]}
\label{fig:small_scale}
\end{figure}

 {The full extension of the disk in total intensity (H band) and polarimetry (J band) is shown in the S/N map in Fig. \ref{fig:SNR}}, while a smaller FOV is displayed in Fig. \ref{fig:small_scale} for IRDIS (H band), IRDIS-DPI (J band), and IFS (YJ band, collapsed image). The S/N map is the ratio between the reduced image and the azimuthal standard deviation of the same image

The  disk of HD\,32297 is relatively bright compared to the stellar residuals. We detect the disk extending to the same distance as observed by HST-NICMOS \citep[][up to ~3.3$''$ from the star]{Schneider2005}. The disk is almost edge-on and the shape appears globally symmetrical in both the northeast (NE) and southwest (SW) sides as seen in both the total intensity and polarimetric images. Some local asymmetries between the NE and SW sides of the disk are discussed in detail in the photometric analysis in Sect. \ref{sec:photom}. 

The IRDIS images feature a concavity at large separations towards the northwest, which becomes significant at a stellocentric distance of about $2''$. This pattern is observed in both the IRDIS BB\_H image and the DPI BB\_J image (as well observed in Fig.\ref{fig:SNR}). This concavity or bow shape was also observed with HST/NICMOS as mentioned in \cite{Schneider2014}, and attributed to the interaction with the ISM of small particles on very eccentric orbits. 

On closer inspection at a smaller scale, inside $1''$ the disk takes a half-elliptical shape indicative of an inclined ring (likely the planetesimal belt), as suggested first by \cite{Boccaletti2012} and \cite{Currie2012}. The high quality of SPHERE images now allows us to resolve this ellipse, with the innermost part of the disk observed at a stellocentric radial distance as close as 0.15$''$  \citep[compared to 0.5-0.6$''$ in][]{Boccaletti2012}. %The associated inclination is about 88\deg. 
We measure the position of the ring ansa at $0.8-0.9''$. The asymmetry with respect to the major axis is reminiscent of forward scattering suggesting that the bright part is the front side of the disk. Interestingly, the S/N map (Fig. \ref{table:obs_log}) built from the IRDIS BB\_H image suggests that the back side of the disk is also detected which is also visible in the IRDIS BB\_H intensity image (Fig.\ref{fig:small_scale}). This is discussed in Sect. \ref{sec:model}. 

Even though the photometry of the disk is impacted by the so-called ADI self-subtraction \citep{Milli2012}, the intensity along the disk varies monotonically, and so does not feature any sign of a gap contrary to the observation reported in \cite{Asensio-Torres2016}. Aditionally, the stellar residual halo in the SPHERE images is much lower than in HiCIAO observations, confidently ruling out such a gap. This is confirmed with our photometric analysis discussed further in Sect. \ref{sec:photom}. 

The IFS image features a highly symmetrical ring on each side of the minor axis, inside the achievable FOV corresponding to stellocentric distances of $<0.9''$. 

In polarimetry, the intensity along the disk varies differently than in total intensity because the scattering phase function is vastly different. As a result, we observe a peak of intensity located at $\sim 0.8''$ and associated to the location of the disk ansa. Due to the absence of self-subtraction in DPI, strong signal is observed at small angles < 0.15$''$ (Fig. \ref{fig:small_scale}). At this stage, it is not yet clear whether this signal is produced by stellar residual or by the disk itself, although it is moderately visible in the S/N map (Fig. \ref{fig:SNR}). There are two possible {explanations} for this signal to be a true disk signal, one being a strong forward scattering peak at small angular separation, and another being a detection of a potential inner belt.

\subsection{Position angle of the disk}
\label{sec:pa}
The position angle (PA) is determined by a method developed for edge-on disks as presented in \cite{Lagrange2012}. The science data are first rotated to an initial guess of the PA, and a Gaussian profile is fitted perpendicular to the disk mid-plane in a range of angular separations from which the spine of the disk and the local slope, globally or separately for the NE and SW sides, are derived. This process is repeated until the slope reaches a minimum corresponding to a horizontal disk and providing a measurement of the actual disk PA. 
The average PA retrieved for both total intensity and polarimetric data is $47.60\degb\pm0.2\degb$ and $47.50\degb\pm0.15\degb$ , respectively. The errors consist of measurement error for the applied method and the additional TN uncertainty of 0.1\deg.

\begin{figure}
\centering
\captionsetup[subfigure]{labelformat=empty}
\subfloat[]{
        \includegraphics[clip, trim={20 0 20 20},width=\hsize]{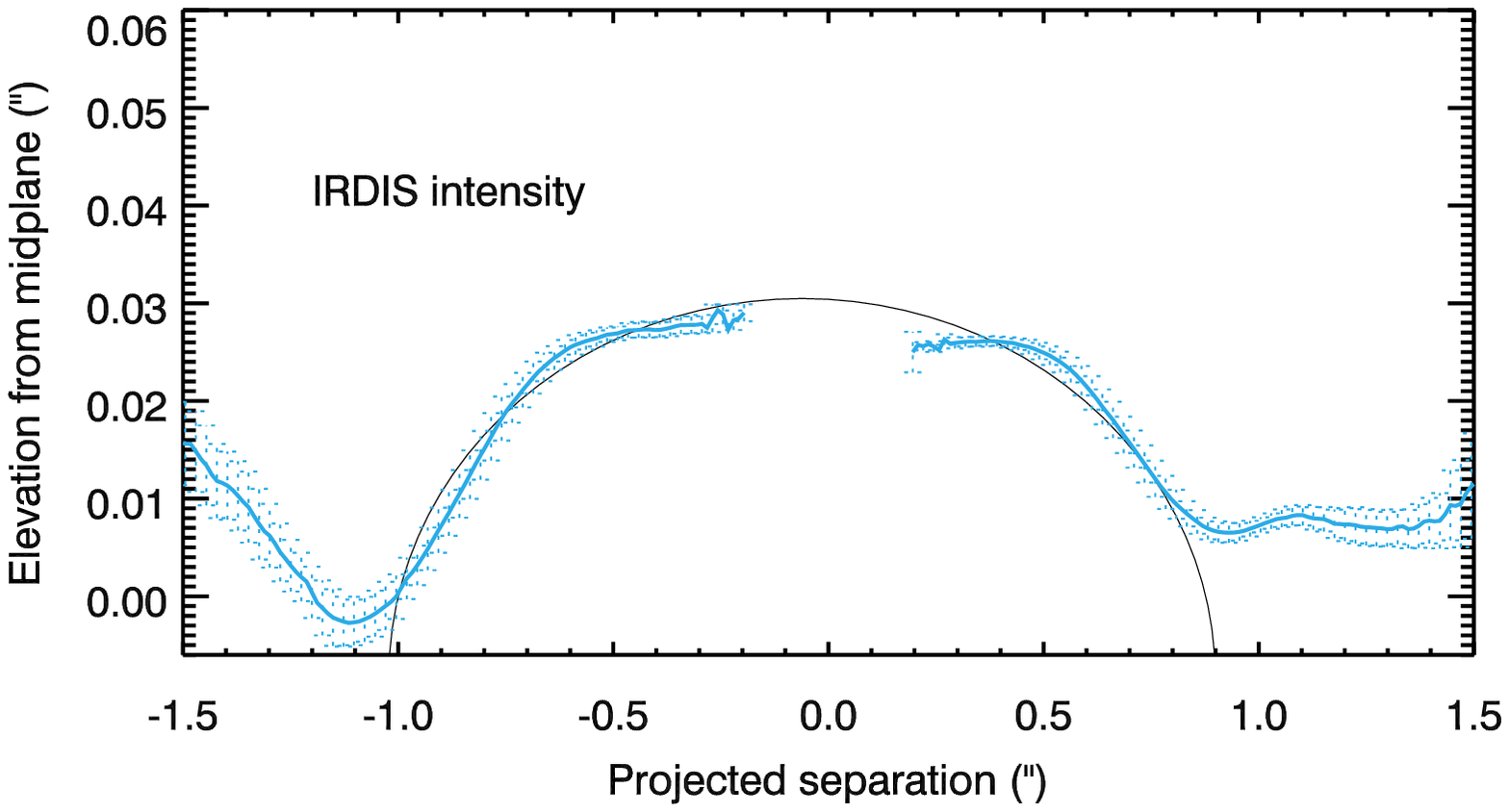}}
\vspace{-25pt}
\subfloat[]{
\includegraphics[clip, trim={20 0 20 0},width=\hsize]{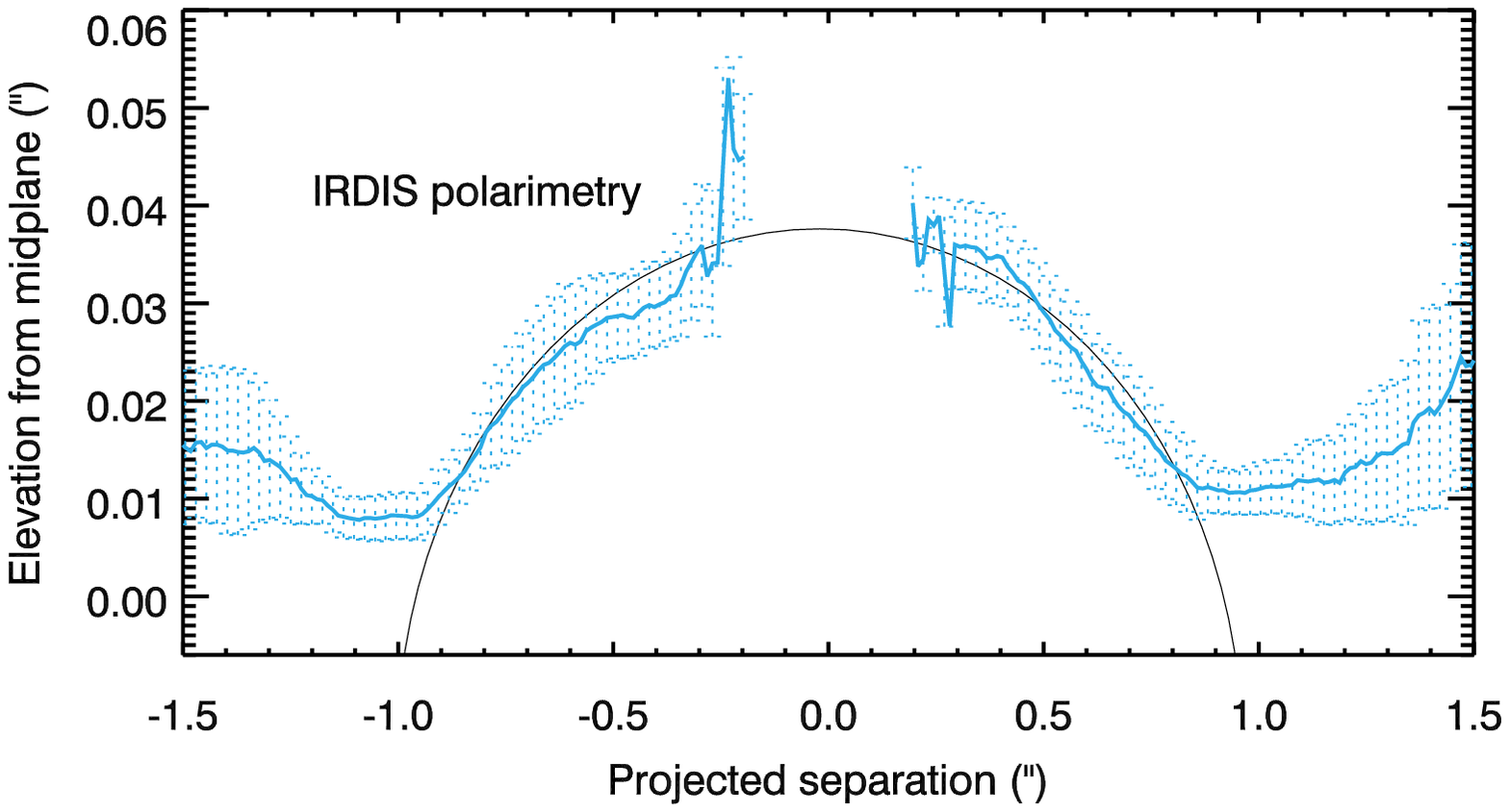}}
\caption{Spine of the disk measured in total intensity image in BB\_H (top) and in polarimetric image in BB\_J (bottom). The spines are fitted with an ellipse.}
\label{fig:spine}
\end{figure}

 The local minima in the spine of the disk can determine the position of the ansae \citep{Mazoyer2014}. For higher accuracy, a simple ellipse fit to the spine of the disk can give the position of the ansae corresponding to the semi-major axes of the ellipse. Additionally, the inclination can be derived from the semi-major and semi-minor axes of the fitted ellipse. The ellipse fitting the total intensity data is centered at (-0.06$''$,-0.008$''$) and  (-0.02$''$,-0.015$''$) for the polarimetric data. The plots are presented in Fig. \ref{fig:spine}. We do this ellipse fit to the spine measured on all the images reduced by the methods,
TLOCI \citep{Marois2015}, KLIP 3,5,10 modes \citep{Soummer2012}, classical ADI \citep{Marois2006ADI}, No ADI \citep{Galicher2018}, and DPI $Q_{\phi}$ \citep{Schmid2006}, and the dispersion between these measurements is used as error. The position of the ansae measured from the spine is $126.4\pm{12.8}$ au and the inclination is $88.4\degb\pm{0.6}\degb$. 

%%%%%%%%%%%%%%%%%%%%%%%%%%%%%%%%%%%%%%%%%%%
%%%%%%%%%%%%%%%%%%%%%%%%%%%%%%%%%%%%%%%%%%%

\section{Geometrical modeling}
\label{sec:model}
\subsection{Modeling total intensity with a single phase function}
\label{sec:model1g}

Angular differential
imaging techniques induce biases to the disk photometry due to self-subtraction. To overcome this problem and recover the unbiased photometry of the disk we proceed with forward modeling, which involves the generation of synthetic images from a model, given some parameters, undergoing a similar post processing to that used on the original data. 
We used the GRaTer model to generate synthetic images of debris disks \citep{Augereau1999}.
The model assumes a ring of planetesimals releasing dust from a collisional cascade and located at  a distance $r$ from the star. The position $R_0$ is where the dust density peaks corresponding to the {position of the ansae} and we then impose that the dust density distribution scales radially as ${\scaleto{r^{\alpha_{in}}}{9pt}}$ inwards and ${\scaleto{r^{\alpha_{out}}}{9pt}}$ outwards. The vertical distribution is fixed to be a Gaussian function, the height of which is controlled by the parameter $H_0$ obtained at a radius  $R_0$.

From this geometrical prescription, GRaTer calculates, for a given model, the resulting scattered light image taking into account the scattering angle $\theta$, which depends on the inclination and PA. The phase function is given by the Henyey Greenstein (HG) function \citep{Henyey1941} for total intensity 

\begin{equation} 
\label{e3}
\begin{aligned}
f_I (\theta) = \dfrac{1-g^2}{4\pi(1+g^2-2g\cos{\theta})^{3/2}} 
\end{aligned}
,\end{equation}

\noindent where $g$ is an anisotropic scattering factor that we leave as a free parameter. For $0<g<1,$ the scattering by dust particles is predominantly forward (isotropic if $g=0$), and conversely backward for  $-1<g<0$.

The free parameters of our model are listed below. The initial guesses are obtained from previous work on this system \citep{Currie2012,Boccaletti2012} and first-order estimations from the SPHERE images, totalling 23040 models. 

\begin{itemize}
    \item  inclination i (\deg): 87.5, 88.0, 88.5, 89.0;
    \item position of the ansae $R_0$ (au): 115, 120, 125, 130,135, 140,145,150;
    \item  power-law index $\alpha_{in} $: 2, 5, 8, 10;
    \item power-law index $\alpha_{out} $:  -4, -5, -6, -7,-8
;    \item HG parameter $g$:  0.4, 0.5, 0.6, 0.7, 0.8, 0.9;
    \item disk aspect ratio $h = H_{0}/R_{0}$ : 0.01, 0.015, 0.02, 0.025, 0.03, 0.035.
\end{itemize}

The PA is kept constant at 47.6\deg\, as a result of the analysis presented in Sect. \ref{sec:pa}.
{Each GRaTer model is first injected into an empty data cube void of noise and convolved with the measured PSF}.
To account for the ADI photometric bias in the forward modeling, a given GRaTer model defined by a set of parameters is projected onto the KL basis which was formed {and used} for the data. Each model is then normalized to the maximum PSF intensity as formerly done for the data (Sect. \ref{sec:obs}). 
Looking for the best-fit model implies that we compare the science image with a series of model images in an appropriate region encompassing the pixels containing disk signal. A numerical mask  $\sim0.15''\times 3''$, aligned with the disk mid-plane, is applied, which is compared to observations in a $\chi^2$ fashion. The central part with a stellocentric distance lower than $\sim0.15''$ is removed from the mask. The northeast and the southwest parts are analyzed separately. {A comparison between the science image and the noiseless model masked with an effective aperture is credible as the disk has high S/N assuming that there is no over-subtraction due to the combination of noise and disk. Also, in \cite{Boccaletti2019} a comparison between forward modeling with noiseless models and the injected model into the science image at a different PA with a certain flux resulted in similar $\chi^2$ values. Therefore, we refrain from doing the latter due to reduced computational momentum.}

The reduced $\chi^2$ is calculated between the science image ($S_{i,j}$) and the models ($M_{i,j}$) at $i,j^{th}$ pixel and summed over the number of pixels in the mask ($N_{data}$). The calculation is described as follows:

\begin{equation} 
\label{eq4}
\begin{aligned}
\chi_\nu^2 =\dfrac{1}{\nu}  \sum_{i,j=1}^{N_{data}}\Bigg( \dfrac{S_{i,j}-a . M_{i,j}(p)}{\sigma_{i,j}}\Bigg)^2,
\end{aligned}
\end{equation}

\noindent where $\nu$ is the degree of freedom equal to $ N_{data} - N_{param}$, where $N_{param}$ is the number of free parameters and $p$ the parameter space explored in these models. 
{The noise term ($\sigma_{i,j}$) is derived from the azimuthal standard deviation in the image masking the disk.} In high-contrast imaging, the noise has a spatial structure related to the stellar residuals, varying with wavelengths, while $\chi^2$ minimization applies to Gaussian errors and linear models. These two conditions are not rigorously met in our forward modeling technique, which imposes some limitations to this approach. 
The parameter $a$ is the scaling factor between the data and the model.
 
\begin{figure}
\includegraphics[width=\hsize]{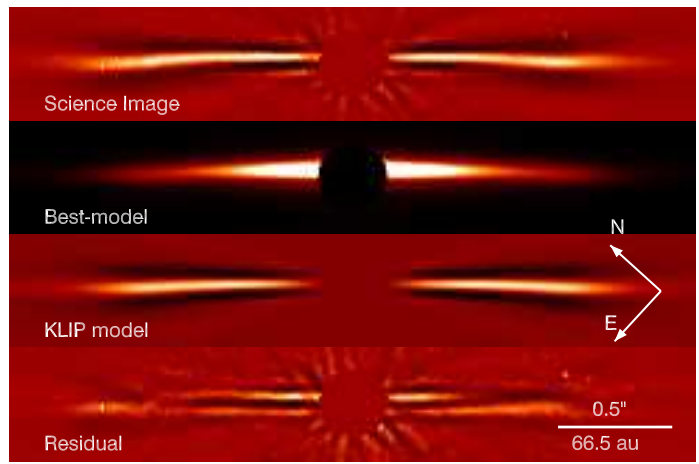}
\caption {From top to bottom: IRDIS science image in BB\_H, 
best GRaTer model with the parameters i=88.5$\degb$, $\alpha_{out}$=-6,{ $\alpha_{in}$=10, $R_0$=130 au}, g=0.60,{ h=0.020}, KLIP processed image for the corresponding GRaTer model, and the residual image. The images are cropped at $3''\times0.5''$ and rotated at 90\deg\,-\,PA.
The science image, KLIP processed image, and the residual image are scaled linearly in the range [$-1\times10^{-5}$,$1\times10^{-5}$] and the GRaTer model is scaled linearly in the range [0.0,0.5] }
\label{fig:int1g}
\end{figure}

To identify the best models, we select 1$\% $ of those with the lowest reduced $\chi^2$ values.
This approach is considered instead of taking models corresponding to the 1$\sigma$ deviation of the reduced $\chi^2$ distribution using the theoretical threshold of $\sqrt{2\nu}$ . This is because there are very few models falling into this latter category, which puts overly strong restrictions on the measurements of error bars. {Also, this threshold is theoretically applicable for Gaussian errors and linear models which, as discussed earlier, our models do not follow.}
As a final output we derive the histogram for each parameter, for the given set of best models. {A Gaussian profile is fitted to the histograms and Table \ref{table:int1g} provides the peak of the Gaussian and the $1\sigma$ deviation from the measured peak as errors. Albiet, in two scenarios a Gaussian cannot be fitted. The first is when the distribution of the histogram is flat. The second is when the number of values explored for the parameter are less than four}. The best-fit model {taken for creating} the residual image as shown in Fig. \ref{fig:int1g} is the model which has the minimum value of the reduced $\chi^2$ (${\chi^{2}=5.26}$).

\begin{table}[h]   

\caption{Parameters that provide the best GRaTer model fitting of the IRDIS BB\_H science image with the one HG phase function} 
\begin{tabular}{c|c c c }     % 7 columns  
\hline \hline  

 & NE side & SW side & Full disk\\
\hline
  i (\deg)  &  88.2$\pm{0.4}$   &{88.3${\pm{0.3}}$} & {88.3${\pm{0.3}}$} (88.5) \\ 
  $R_0$ (au) &{ 136.8${\pm{10.0}}$}  &{129.4${\pm{8.4}}$}& {134.7${\pm{9.3}}$ (130)}\\
  $\alpha_{in} $ &{ 6.00${\pm{4.00}}$} &{ 6.00${\pm{4.00}}$} & {6.00${\pm{4.00}}$ (10)}\\
  $\alpha_{out}$ & {-6.0${\pm{0.9}}$} & -5.7$\pm{0.8}$ & {-6.17${\pm{1.17}}$} (-6)\\
  $g$ & {0.55${\pm{0.14}}$ }&{0.49${\pm{0.14}}$} & 0.55$\pm{0.13}$ (0.6)\\
  $h$ &{ 0.026${\pm{0.005}}$} &0.024$\pm{0.006}$& {0.020${\pm{0.006}}$ (0.020)}\\
\hline \hline                 
\end{tabular}
\tablefoot{The model parameters corresponding to the smallest $\chi^{2}$ value are provided in brackets in the fourth column. {The mean value of the distribution and the dispersion is used for $\alpha_{in}$ as its histogram has a flat distribution}}
\label{table:int1g} 
\end{table} 

The inclination we find,{ $88.3\degb\pm0.3\degb$}, is compatible with previous measurements \citep{Boccaletti2012,Currie2012} but achieves a higher accuracy. The {position of the ansae $R_0$ is ${134.7\pm9.3}$ au}. 
The variation of the dust density inwards from the belt ($\alpha_{in}$)
is relatively difficult to constrain for inclined disks; only limited values are explored for this parameter and evidently the retrieved value of ${6.0\pm4.0}$ is not constrained.{ On the other hand, the values for $\alpha_{out}$, $g,$ and $h$ are relatively well constrained with values of ${-6.17\pm1.17}$, $0.55\pm0.13,$ and ${0.020\pm0.006}$, respectively}.

From Fig. \ref{fig:int1g}, one can see that the residuals are still relatively large, which indicates that the adopted model {does not perfectly} explain the data. Using two HG phase functions instead of one has been proved effective to better fit inclined disks \citep{Currie2012,Milli2017}. Similarly, we also try to improve our model by using two HG phase functions to model the disk.  

\subsection{Modeling total intensity with two HG phase functions}
\label{sec:model2g}

{The previous section uses a single HG function to model the scattering phase function but allowed to explore a large range if parameters}. In the total intensity IRDIS image, a faint signature of the back side is visible (Fig.~\ref{fig:SNR}). In trying to better model the back scattered grains in the disk, we adopt a new phase function based on two HG functions as shown below.

\begin{equation} 
\label{eq5}
\begin{aligned}
f_T (\theta) = w_1.f_I (g_1,\theta) + (1-w_1).f_I (g_2,\theta)
\end{aligned}
,\end{equation}

\noindent where $g_2$ is assumed to be negative as it models the backward scattering component of the disk.

{We reduced the parameter space guided by our results from the previous section and thus the explored parameters for this case are}:

\begin{itemize}
    \item inclination i (\deg): 88.0, 88.5;
    \item {position of the ansae} $R_0$ (au):120, 125, 130, 135, 140, 145;    \item power-law index $\alpha_{in} $: 2, 5, 8, 10;
    \item power-law index $\alpha_{out} $:  -5, -6, -7;      
    \item first HG parameter g$_1$:  0.4, 0.5, 0.6, 0.7;
    \item second HG parameter g$_2$:-0.5, -0.4, -0.3, -0.2, -0.1;
    \item disk aspect ratio  $h$ : 0.02, 0.025, 0.03;
    \item weight $w_1$: 0.80, 0.83, 0.87, 0.90, 0.93.
\end{itemize}

The range of weights $w_1$ is chosen to encompass the values derived for \citep[HD\,32297 with KECK by][$w_1=0.90$]{Currie2012} and another inclined debris disk \citep[HR\,4796][$w_1$=0.83]{Milli2017}.

With the combination of the above listed parameters, 43200 models are created, for which the reduced $\chi^2$ is measured and the best models are identified following the same approach as described in Sect.~\ref{sec:model1g}. 
The parameters giving the best-fit model for this case are provided in Table~\ref{table:int2g}.

\begin{table}[h]
\caption{Parameters that provide the best GRaTer models fitting the IRDIS BB\_H KLIP science image with two HG phase functions} 
\begin{tabular}{c | c c c  }     % 7 columns  
\hline \hline  
  & NE side & SW side & Full disk\\
\hline
  i (\deg)  & {88.0}   &88.2${\pm{0.2}}$ & {88.2${\pm{0.2}}$} (88.0) \\ 
  $R_0$ (au) & {132.4${\pm{8.3}}$  } & {128.5${\pm{8.6}}$ }& {134.4${\pm{8.5}}$ }(140)\\
  $\alpha_{in} $& {5.2${\pm{2.8}}$  } & 6.0$\pm{4.0}$ & {6.0${\pm{4.0}}$ }(2)\\
  $\alpha_{out}$& {-6.0${\pm{1.0}}$  } &{ -6.0${\pm{1.0}}$ }&{ -6.0${\pm{1.0}}$ }(-6)\\
  $g_1$ &{ 0.68${\pm{0.05}}$  } & 0.69$\pm{0.06}$ &{ 0.69${\pm{0.06}}$ }(0.7)\\
  $g_2$ & {-0.3${\pm{0.2}}$  } &{-0.3${\pm{0.2}}$ }& {-0.3${\pm{0.2}}$ }(-0.4)\\
  $w_1$ & 0.82$\pm{0.04}$   & {0.80${\pm{0.06}}$ }&{ 0.81${\pm{0.05}}$ }(0.80)\\
  $h$ & {0.025${\pm{0.005}}$ }  &0.022$\pm{0.003}$ &{ 0.022${\pm{0.002}}$ }(0.020)\\
\hline \hline                 
\end{tabular}
\tablefoot{The model parameters corresponding to least $\chi^{2}$ are provided in bracket in the fourth coloumn.{The mean value of the distribution and the dispersion is used for i, $\alpha_{in}$, $\alpha_{out}$,$g_2$ and h as they have either flat distribution or small parameter space.}}
\label{table:int2g} 
\end{table}

The inclination is very well constrained to ${88.2\pm{0.2}}$\deg, but we highlight the fact that we considered only two values to explore the parameter space as this is relatively well-constrained from previous studies as well as in the previous section (Sect.~\ref{sec:model1g}).  The {position of the ansae and the inner power-law index are found to be ${R_0=134.4\pm{8.5}}$ au and ${\alpha_{in}=6.0\pm{4.0}}$, and ${\alpha_{out}}$ is ${-6.0\pm{1.0}}$ which are within the error bars retrieved in the previous section. The value of ${h = 0.022\pm{0.002}}$ is also consistent with results from the previous section.} 

{We find ${g_1 = 0.69\pm{0.06}}$, ${g_2=-0.3\pm{0.2,}}$ and ${w_{1} = 0.81\pm{0.05}}$.} These values do not agree with the best fits in K band in \cite{Currie2012}, where a similar HG function to Eq. \eqref{eq5} was used to model the phase component with $g_1 = 0.96$, $g_2=-0.1,$ and $w_{1} = 0.9$. It should however be considered that our data have better S/N compared to the observations of \cite{Currie2012} and we observe the back side of the disk in our data.

\begin{figure}
        \includegraphics[width=\hsize]{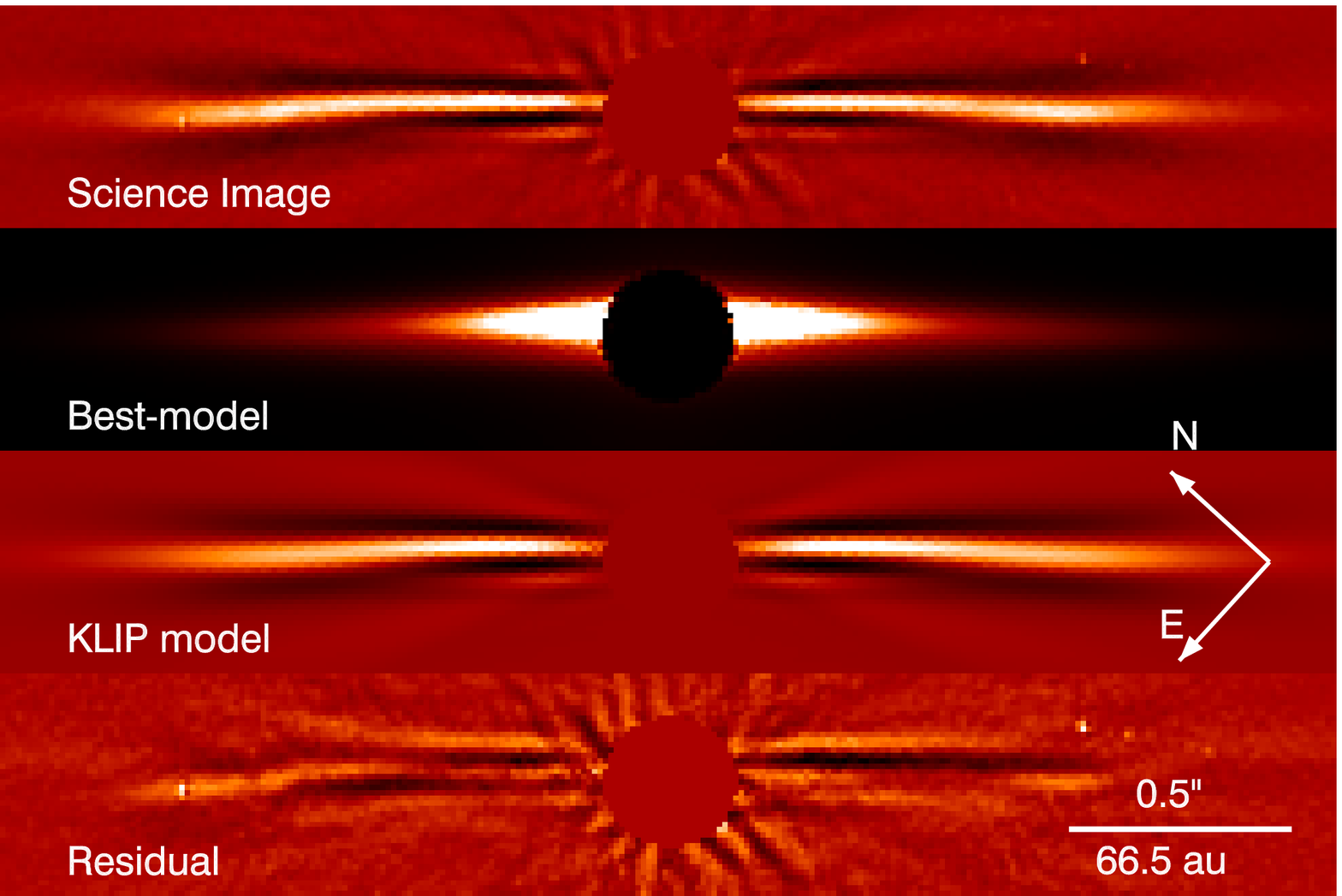}
\caption{ (a) From top to bottom: IRDIS science image in BB\_H, best GRaTer model with the parameters  i=88$\degb$, $\alpha_{out}$=-6, $\alpha_{in}$=2, $R_0$=140 au, $g_1$=0.70, $g_2$=-0.4, $w_1$=0.80, h=0.020, KLIP processed image for the corresponding GRaTer model, and the residual image. All the images are cropped at $3''\times0.5''$ and rotated to 90\deg\,-\,PA.
The science image, KLIP processed image, and the residual image are scaled linearly in the range [$-1\times10^{-5}$,$1\times10^{-5}$] and the GRaTer model is scaled linearly  in the range [0.0,0.25]}
\label{fig:int2g}
\end{figure}

In Fig.~\ref{fig:int2g}, we plot our {best-fit model} together with the data and the residuals found after subtracting them both. {The reduced $\chi^2$ value for the best-fit model is 3.29}. We find that our model provides a 31\% better match to the data compared to models with single HG phase function. Irrespective of the improvement,  some residuals remain, which may be due to limitations that are combination of several factors. The simplicity of our model with only eight parameters cannot reproduce the complexity of the disk; the variation of the PSF during observation, the possibility of over-subtraction and indirect self-subtraction in the forward modeling \citep{Pueyo2016}, along with nonlinear terms ignored in this study can all add to limitation of our best-fit model.

Independently fitting the disk observed in each spectral channel of IFS with the same set of parameters ($R_0$, $\alpha_{in}$, $\alpha_{out}$, h)  but $g_1$, $g_2,$ and $w_1$ being free parameters, we found that the variation of the anisotropic scattering factors is no larger than 8\%. This gives confidence in selecting the same value for all spectral channels. We note that the disk is observed only partly in the IFS FOV, and therefore any degeneracy between morphological and phase parameters is not considered in this test. 

\subsection{Modeling polarimetric images}
\label{sec:model_pol}

Polarimetric observations complement total intensity data and depend differently on the  morphological parameters. 
For instance, as shown in \citet{Engler2017}, the phase function in polarimetry peaks close to $\sim90\degb$ phase angle as opposed to the total intensity phase function, which reaches a maximum at small phase angles. As a result, the ansae has stronger signatures in polarimetry than in total intensity. 
Modeling the polarimetric image is therefore crucial to derive the geometry of the debris disk \citep{Olofsson2016}. 
 
 We use GRaTer to create a grid of 32256 geometrical models with a polarised phase function $f_P (\theta),$ given below:

\begin{equation} 
\label{eq6}
\begin{aligned}
f_{P}(\theta) = f_{I}(\theta)\dfrac{1-\cos^2 \theta}{1+\cos^2 \theta}
\end{aligned}
,\end{equation}
 
\noindent where $\theta$ is the scattering angle. The phase function used is the combination of the HG function (Eq.~\ref{eq5}) and Rayleigh scattering, to account for the angular dependence of linear polarization due to single scattering of an optically thin disk.

The back side of the disk is not visible in the polarimetric image, and therefore we used a single HG function to construct the phase function in the GRaTer models.  The free parameters are as follows:

\begin{itemize}
    \item  inclination i (\deg): 87.5, 88.0, 88.5, 89.0;
    \item {position of the ansae} $R_0$ (au): 115, 120, 125, 130,135, 140,145,150;
    \item  power-law index $\alpha_{in} $: 2, 5, 8, 10;
    \item power-law index $\alpha_{out} $: -3,  -4, -5, -6, -7,-8;
    \item HG parameter $g$:  0.4, 0.5, 0.6, 0.7, 0.8, 0.9, 0.99;
    \item disk aspect ratio $h$: 0.01, 0.015, 0.02, 0.025, 0.03, 0.035.
\end{itemize}

We applied the same procedure as in Sect. \ref{sec:model1g} to measure the reduced $\chi^2$ and derive the range of best models. The value of the reduced $\chi^2$ is { 3.52 for the best-fit model}.

\begin{table}[h]  

\caption{Parameters that provide the best GRaTer model fitting of the IRDIS BB\_J science image} 
\begin{tabular}{c|c c c  }     % 7 columns  
\hline \hline  

  &NE side&SW side&Full disk\\
\hline
   i (\deg)  &88.7$\pm{0.3}$ & 88.5$\pm{0.3}$&  88.6$\pm{0.3}$ (88.5) \\ 
  $R_0$ (au) & {135.8${\pm{9.8}}$}& {127.3${\pm{9.1}}$}& {127.9${\pm{8.0}}$ (125)}\\
  $\alpha_{in} $ & {7.4${\pm{3.2}}$}&{8.1${\pm{3.1}}$ }&  8.1$\pm{3.2}$ (10)\\
  $\alpha_{out}$ & -4.1$\pm{1.0}$&{ -3.7${\pm{0.7}}$}& {-3.9${\pm{0.8}}$} (-4)\\
  $g$ & {0.85${\pm{0.09}}$ }&0.88$\pm{0.05}$ &{ 0.84${\pm{0.08}}$ (0.8)}\\
  $h$ &0.023$\pm{0.006}$ & 0.022$\pm{0.006}$& {0.022 ${\pm{0.006}}$ (0.020)}\\
\hline \hline                 
\end{tabular}
\tablefoot{The model parameters corresponding to the smallest $\chi^{2}$ are provided in bracket in the fourth column}
\label{table:pol1g} 
\end{table}

The parameters for the best-fit model of the polarimetric image are given in Table \ref{table:pol1g}. 
We find that the inclination of the disk is 88.6\deg$\pm{0.3}$\deg and the {position of the ansae} (${127.9\pm{8.0}}$ au) is consistent with the one obtained in total intensity within error bars.
Regarding the slopes of the surface density profile, polarimetry favors a steep inner edge ($\alpha_{in} =8.1\pm{3.2}$) while this parameter was essentially unconstrained in total intensity. {On the contrary, the outer slope ${\alpha_{out}=-3.9\pm{0.8}}$ is flatter}.

\begin{figure}
    \includegraphics[width=\hsize]{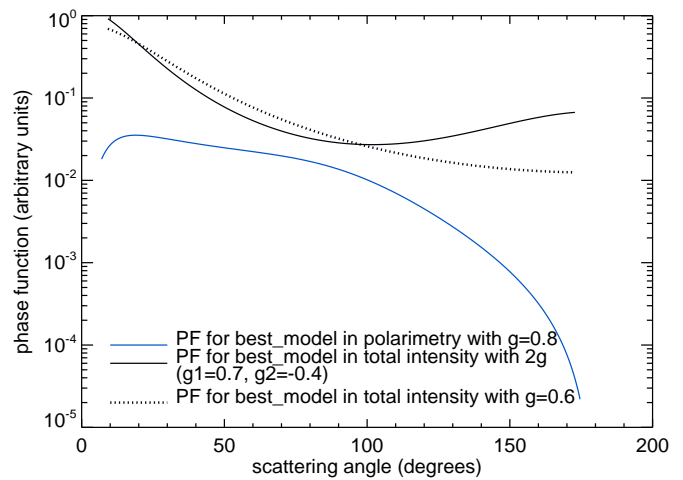}
    \caption{Plots of scattering phase function adopted for the total intensity and polarimetric best-fit models.}
    \label{fig:HG_PF}
\end{figure}

\begin{figure}
\includegraphics[clip, trim={0 0 0 0},width=\hsize]{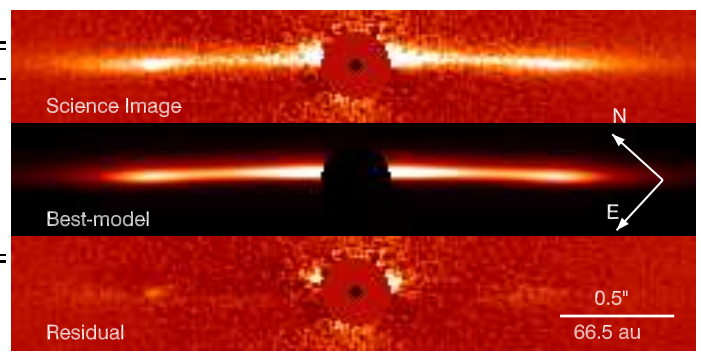}
\caption {From top to bottom: DPI $Q_\phi$ science image, best GRaTer model with parameters i=$88.5\degb$, ${R_0=125au}$, $\alpha_{in}$= 10.0, $\alpha_{out}$=-4.0, {g=0.8, h=0.020,} and the residual image. The images are cropped at $3''\times0.5''$ and rotated at 90\deg\,-\,PA. All images are scaled linearly in the range [$-1\times10^{-5}$, $1\times10^{-5}$]}
\label{fig:pol1g}
\end{figure}

The anisotropic scattering factor is significantly larger in polarimetry ${(0.84\pm{0.08}})$ as compared to intensity irrespective of the number of $g$ parameters we considered. Therefore, the dust grains which are probed in these data are more prone to forward scattering in polarimetry. It should be noted that the in case of strong forward scattering, the polarised phase function peaks at a smaller scattering angle as well as at the ansae \citep{Milli2019}, which is visible in our best-fit model as seen in Fig. \ref{fig:pol1g}. {Figure \ref{fig:HG_PF} plots the scattering phase functions for total intensity and polarimetric best-fit models. The scattering angles probed for HD\,32297 are between 6\deg\ and 175\deg. The phase function in polarimetry peaks at scattering angles smaller than $\sim90\degb$ as seen in Fig.~\ref{fig:HG_PF} while in total intensity we observe an increase beyond 110$\degb$ due to the back-scattering for a double HG function.}

Finally, although the residuals displayed in Fig. \ref{fig:pol1g} are much lower than for the modeling of total intensity, there is still some intensity left near the ansae, indicating that the model does not perfectly reproduce the disk. The residuals could also be an indication of our preference of larger $g$ values over a possibility of an inner component at a separation of < 40 AU. Modeling the polarimetric observation with consecutive inner and outer belts is beyond the scope of this paper 

\section{Photometry and analysis}
\label{sec:photom}
 \subsection{Self-subtraction profile and photometry for the total intensity}
\label{sec:photom_inten}

There are several possible approaches to retrieving the spectrophotometry of the disk. We compare two methods which we use to retrieve the surface brightness of the disk and discuss the limitations associated with each of them.

\begin{itemize}
\item\textit{Method 1}:
The first method is similar to the one used to measure the photometry of the HD\,32297 disk in \cite{Boccaletti2012} and HD\,15115 in \citet{Mazoyer2014}. It was developed for highly inclined debris disks and relies on estimating the self-subtraction caused by the ADI process.
The edge-on geometry allows simplification of this calculation to a 1D problem.
Given the best fit model fitting the data identified in Sect. \ref{sec:model2g}, we first extracted the radial profile of the GRaTer model (convolved with the PSF) and its associated KLIP image. The profiles for both the model and the KLIP image are measured in the same numerical mask as for the $\chi^2$ minimization, and we average the flux in nonoverlapping concentric arcs of $0.15''$ vertical width, and four-pixel ($\sim0.05''$) length.

The self-subtraction is derived from the ratio of these two profiles (Fig. \ref{fig:selfsub}, left) and can be significant at short separations when using KLIP (about 60 at $0.2''$ in the H band). 
The unbiased surface brightness (in counts) of the disk is then obtained by multiplying the self-subtraction profile with the science (KLIP) image profile to compensate for the ADI effect, in each spectral channel of the IFS and each filter of IRDIS. 

Self-subtraction could also be deduced directly from the ratio of the images of the GRaTer model (PSF convolved) and its KLIP version instead of using profiles. \citet{Milli2017} used this method in the case of a less inclined disk and using a specific ADI process (masked classical ADI), which minimizes the self-subtraction
beforehand. Here, in the case of a highly inclined disk together with KLIP processing, this method would lead to the self-subtraction profile presented in the right panel of Fig. \ref{fig:selfsub}, which clearly cannot be used for photometric correction. The large variations are attributed to the strong positive to negative fluctuations resulting from the KLIP processing. Therefore, we conclude that this method of self-subtraction measurement should be avoided for KLIP-processed data.

\vspace{0.5cm}

\item \textit{Method 2}:  The second solution does not require evaluation of the self-subtraction, but instead uses one step of the modeling when the reduced $\chi^2$ is calculated. For the minimum value of reduced $\chi^2$, the scaling parameter $a$ in Eq. \ref{eq4} directly provides the scaling factor between the KLIP image of the best-fit model and the data.  Therefore, contrary to method 1, the surface brightness profile is evaluated in the scaled GRaTer model (PSF convolved) image instead of the real disk image. 

\end{itemize}

For further calculation  we first obtain the stellar flux by integrating over a masked PSF which contains 99.99\% of its total flux. The radius obtained for the mask is 0.4$''$ for the PSF obtained by IRDIS and 0.3$''$ for IFS.
Surface brightness profiles are then converted to magnitude arcsec$^{-2}$ taking into account the pixel size and normalizing  with respect to the stellar flux. 
The error bars linearly combine two terms, the dispersion of the
stellar+background residual intensity as measured in the mask rotated by 90$\degb$ relative to the real disk for each radial bin, and the accuracy of the photometric extraction. The latter is estimated with a fake disk (same parameters as the best-fit model) injected into the raw data at 90$\degb$ from the real disk at a contrast level of $5\times10^{-4}$ and processed in the same way. The photometric extraction is found to be consistent with this initial contrast within $\sim3.3\%$.

{The two methods provide very similar results as seen in Fig. \ref{fig:SB}.}
Hence, we derived surface brightness profiles for all spectral channels of IFS and IRDIS.  {Flowcharts representing the calculations using either method are shown in Fig \ref{fig:FCMethod2}.}
\tikzstyle{proc} = [rectangle, rounded corners, minimum width=3cm, minimum height=1cm,text centered, draw=black, fill=blue!10]
\tikzstyle{startend} = [rectangle, rounded corners, minimum width=3cm, minimum height=1cm,text centered, draw=black, fill=red!10]
\tikzstyle{arrow} = [thick,->,>=stealth]

\vspace{0.5cm}

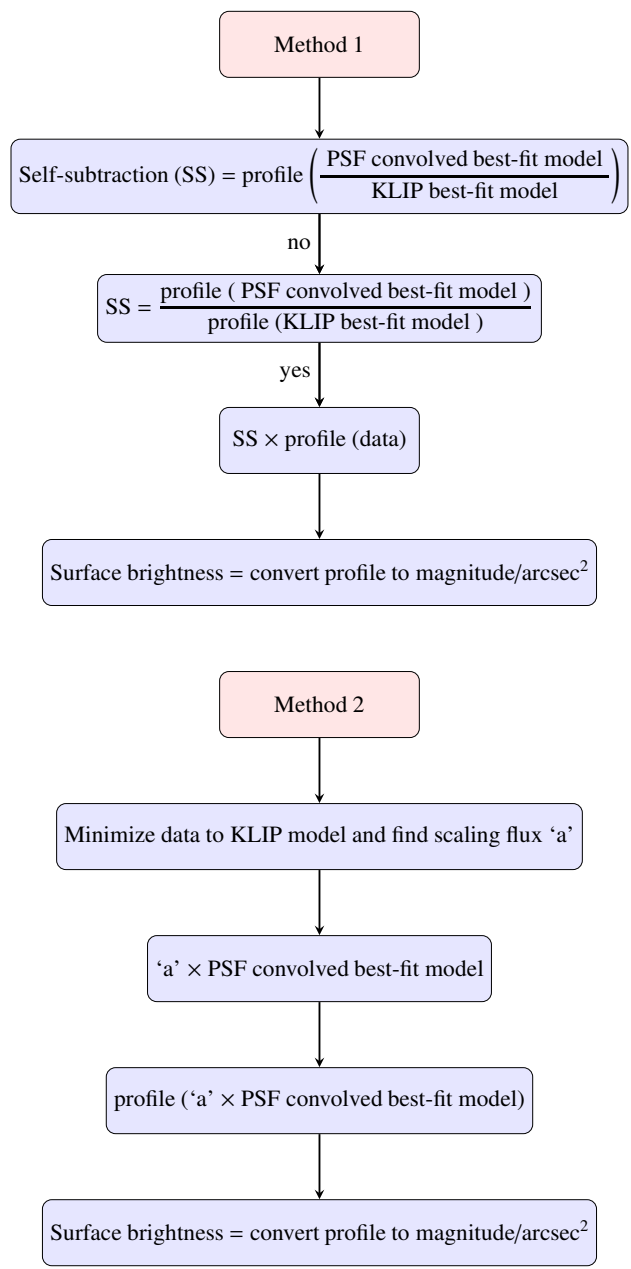
\begin{figure}[!h]
\resizebox {0.45\textwidth} {!} {
\begin{tikzpicture}[node distance=2cm]
\node (o) [startend] {Method 1};
\node (o1) [proc,below of=o] {Self-subtraction (SS) = profile \Bigg($\dfrac{\textrm{  PSF convolved best-fit model }}{\textrm{ KLIP best-fit model }}$\Bigg)};
\node (o2) [proc, below of=o1] {SS = $\dfrac{\textrm{profile ( PSF convolved best-fit model )}}{\textrm{profile (KLIP best-fit model )}}$};
\node (o3) [proc, below of=o2] {SS $\times$ profile (data) };
\node (o4) [proc, below of=o3] {Surface brightness  = convert profile to magnitude/arcsec$^2$};
\node (p) [startend, below of=o4] {Method 2};
\node (p1) [proc, below of=p] {Minimize data to KLIP model and find scaling flux `a'};
\node (p2) [proc, below of=p1] {`a' $\times$ PSF convolved best-fit model};
\node (p3) [proc, below of=p2] {profile (`a' $\times$ PSF convolved best-fit model)};
\node (p4) [proc, below of=p3] {Surface brightness = convert profile to magnitude/arcsec$^2$};
\draw [arrow] (o) -- (o1);
\draw [arrow] (o1) -- (o2);
\draw [arrow] (o2) -- (o3);
\draw [arrow] (o3) -- (o4);
\draw [arrow] (o1) -- node[anchor=east] {no} (o2);
\draw [arrow] (o2) -- node[anchor=east] {yes} (o3);
\draw [arrow] (p) -- (p1);
\draw [arrow] (p1) -- (p2);
\draw [arrow] (p2) -- (p3);
\draw [arrow] (p3) -- (p4);
\end{tikzpicture}
}
\caption{Flowchart representing surface brightness calculations using method 1 or method 2. In Method 1 the arrow anchored with `no' depicts that the calculation of SS with the previous process does not work, resulting in Fig.~\ref{fig:selfsub} (Right). Next, we proceed with another process providing Fig.~\ref{fig:selfsub} (Left) which is used further and therefore the arrow is anchored with a `yes'.} 
    \label{fig:FCMethod2}
\end{figure}

\begin{figure*}[h]
\includegraphics[width=9cm]{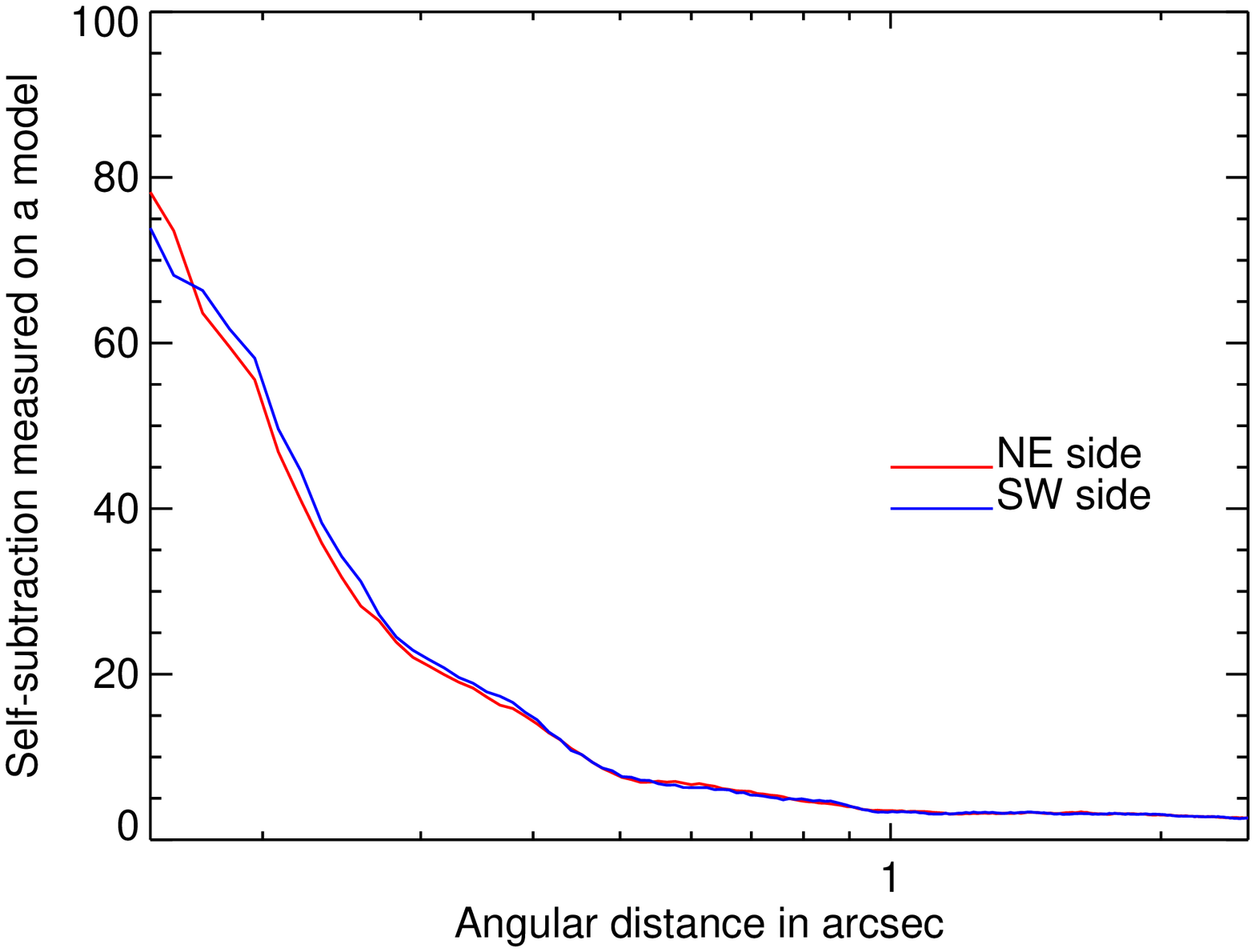}  
\includegraphics[width=9cm]{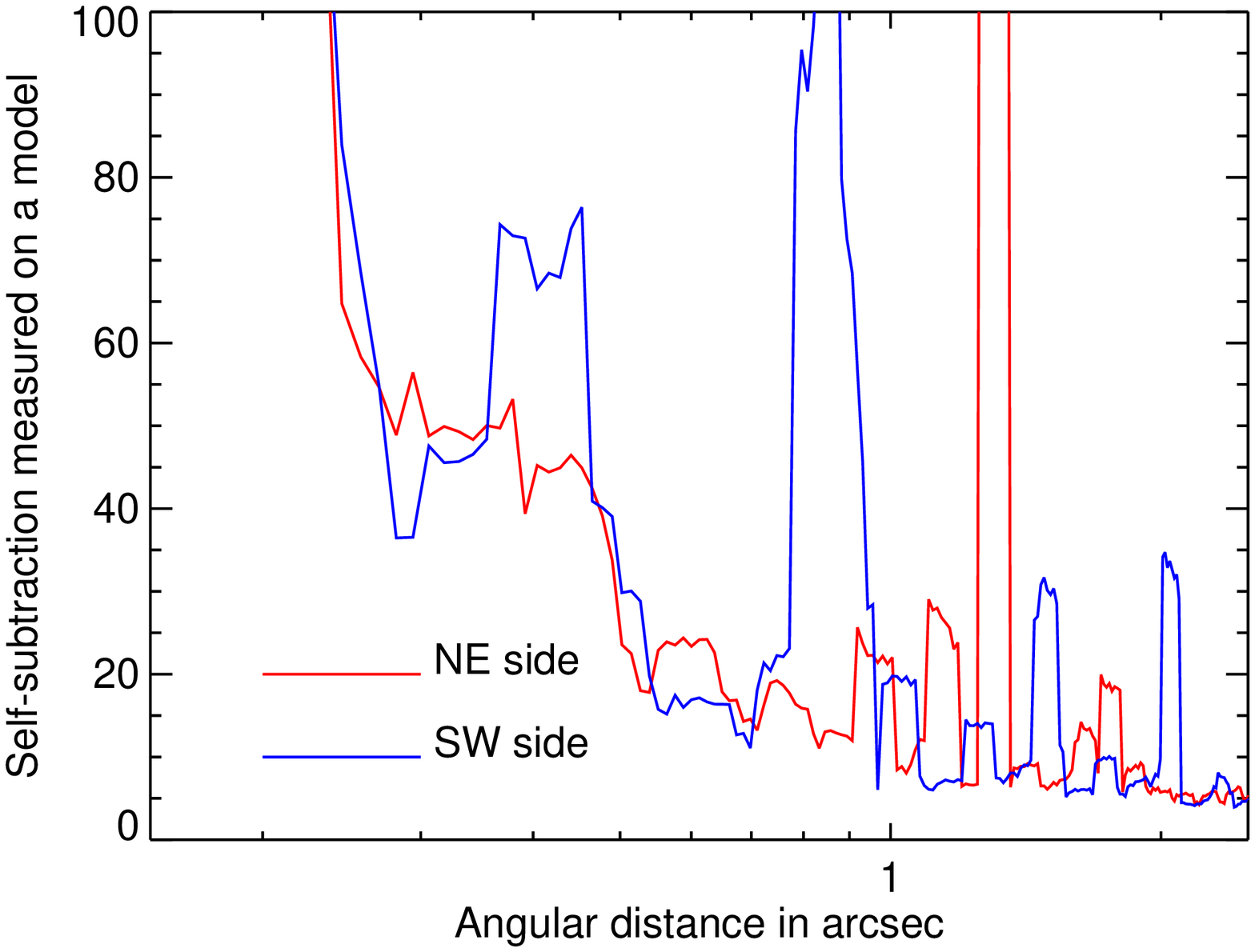}
\caption{ 
Self-subtraction measured for IRDIS in the BB\_H filter for the method 1 (see text for details) using the ratio of radial profiles (left) and the ratio of images (right).}
\label{fig:selfsub}
\end{figure*}

\begin{figure*}[h]
\includegraphics[width=9cm]{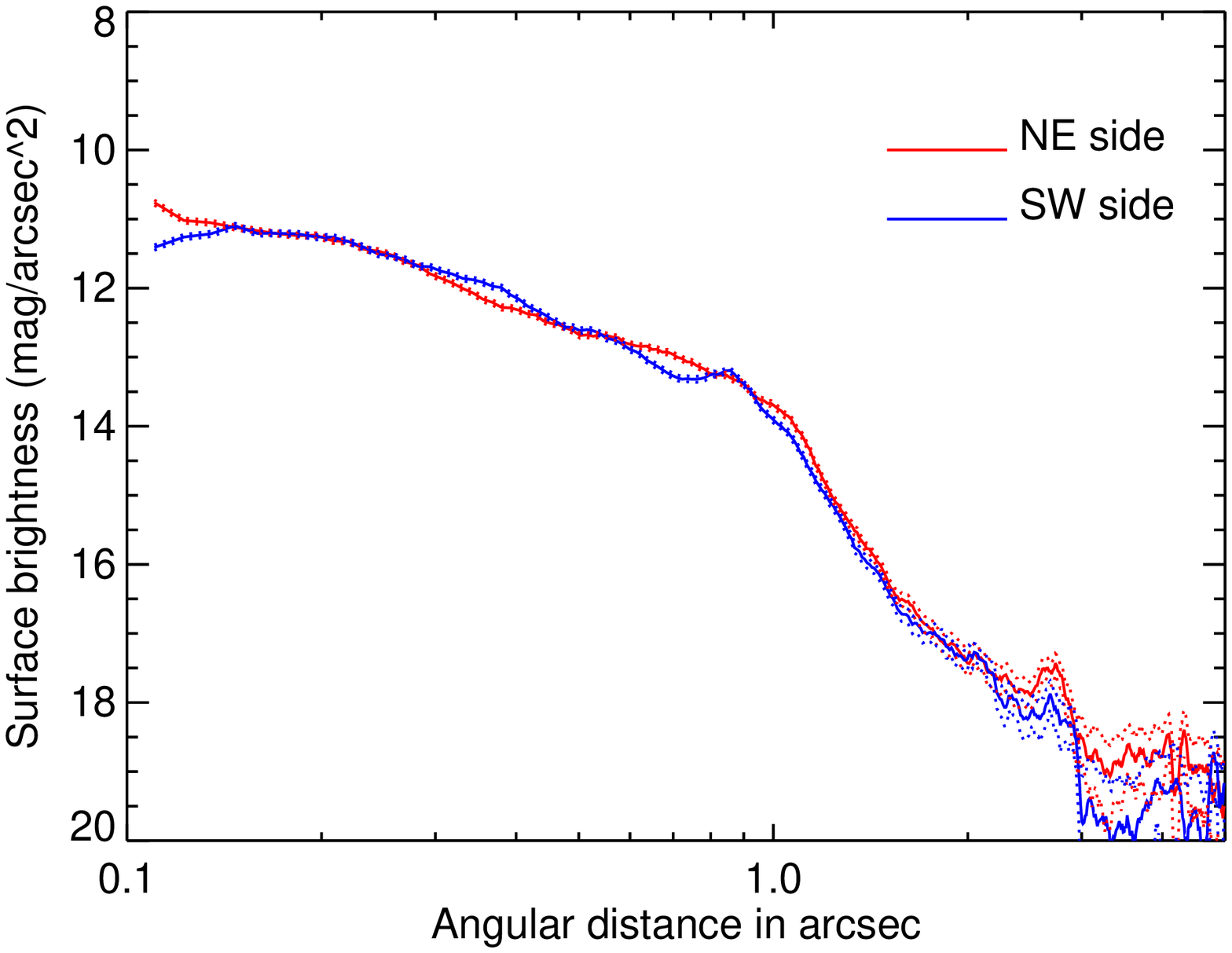} 
\includegraphics[width=9cm]{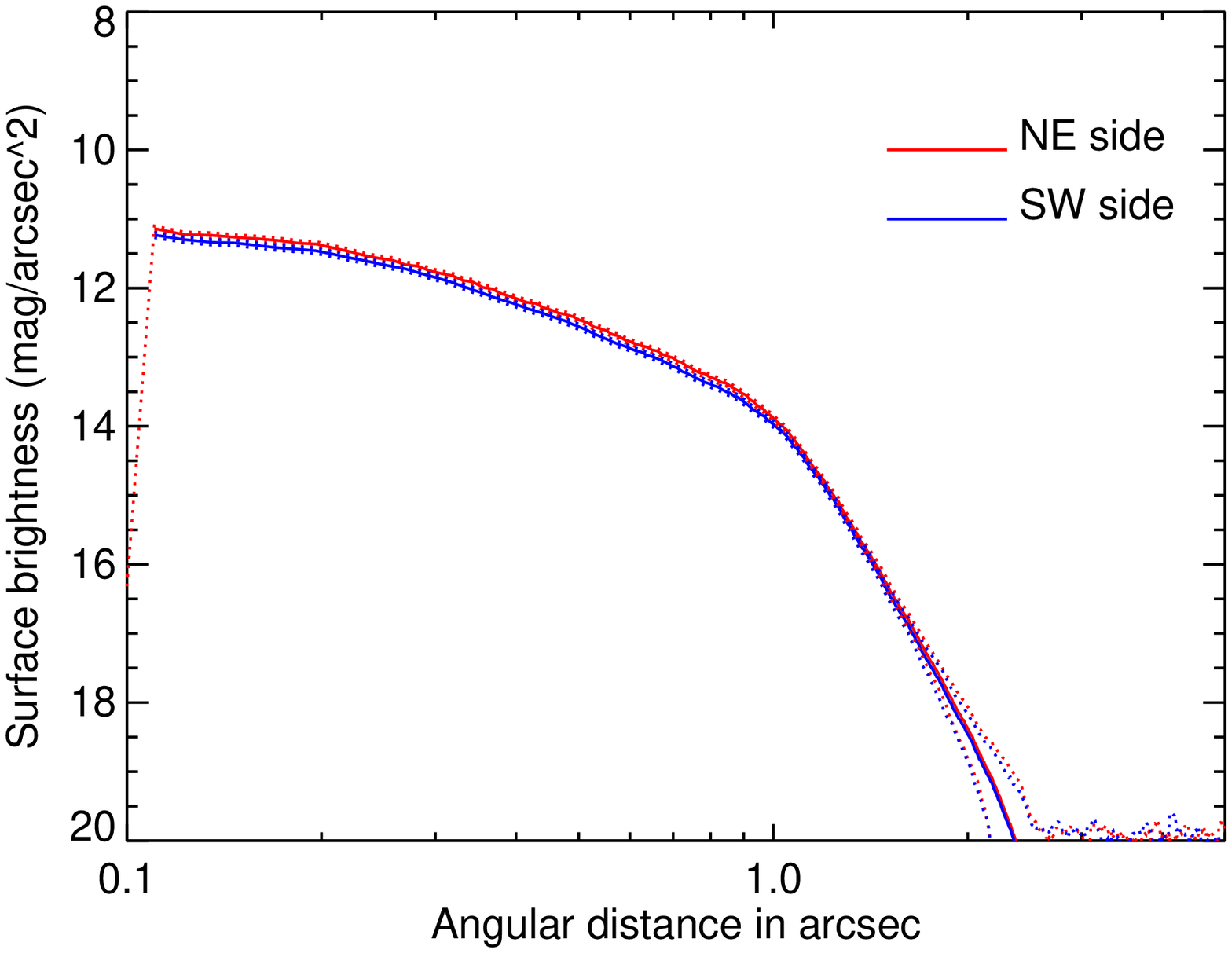}
\caption{ 
Surface brightness of the disk measured for IRDIS in the BB\_H filter for method 1 (left) and method 2 (right). The blue (resp. red) solid line represents the SW (resp. NE) side of the disk. The dotted lines show the errors in the measurements. 
}
\label{fig:SB}
\end{figure*}

At this stage, and contrary to \cite{Asensio-Torres2016}, we do not confirm the presence of a dip or break in the surface brightness profiles near 0.75$''$ on the NE side or 0.65$''$ in the SW,  and therefore we rule  out the presence of a gap (Fig. \ref{fig:SB}). 

\subsection{Photometry for polarimetric images }
\label{sec:photom_pol}

In polarimetry, the disk photometry is not affected by self-subtraction and therefore can be directly measured from the image.
Contrast is calculated directly from the science DPI BB\_J image, instead
of a model, using the same process as explained in method 2 of Sect. \ref{sec:photom_inten}. 

\begin{figure}[h]
\includegraphics[width=\hsize]{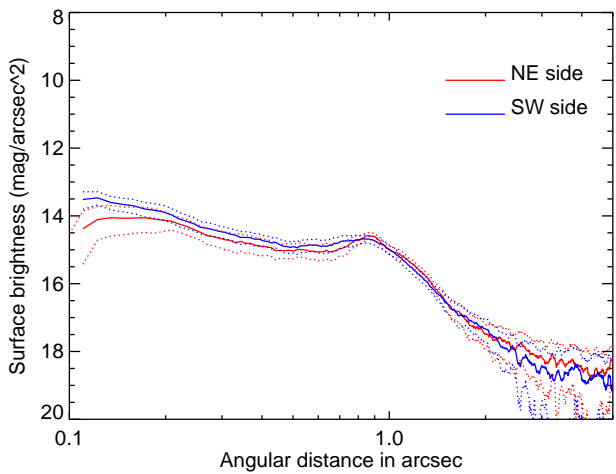}
\caption {Surface brightness profile of the disk measured in the IRDIS BB\_J DPI $Q_\phi$ science image. The blue line represents the SW side of the disk and the red line represents its counterpart on the NE side. The dotted lines represent the error bars. }
\label{fig:SBpola}
\end{figure}

As a result of the scattering angle dependence (Sect. \ref{sec:model_pol}), the slope of the surface brightness profile is clearly different in polarimetry compared to total intensity, with a less steep decrease from 0.1$''$ to about $0.7''$ (Fig. \ref{fig:SBpola}). At the location of the ansae ($\sim0.9''$), the surface brightness shows a peak instead of a break. As advocated in \cite{Engler2018} for the case of the inclined debris disk around HD\,15115, this dependence of the phase function provides a better sensitivity for polarimetric data to pinpoint the inner edge or the ansae of a dust belt, or to reveal multiple structures. In the case of HD\,32297, the polarimetric surface brightness does not show any particular signs of such multiple belts. The polarimetric surface brightness is about two magnitudes fainter than in total intensity for a stellocentric distance of 0.5$''$, which translates to a polarized fraction of about 15\%. 
 
\subsection{Average spectral reflectance}
\label{sec:spect}
\begin{figure*}
\includegraphics[width=9cm]{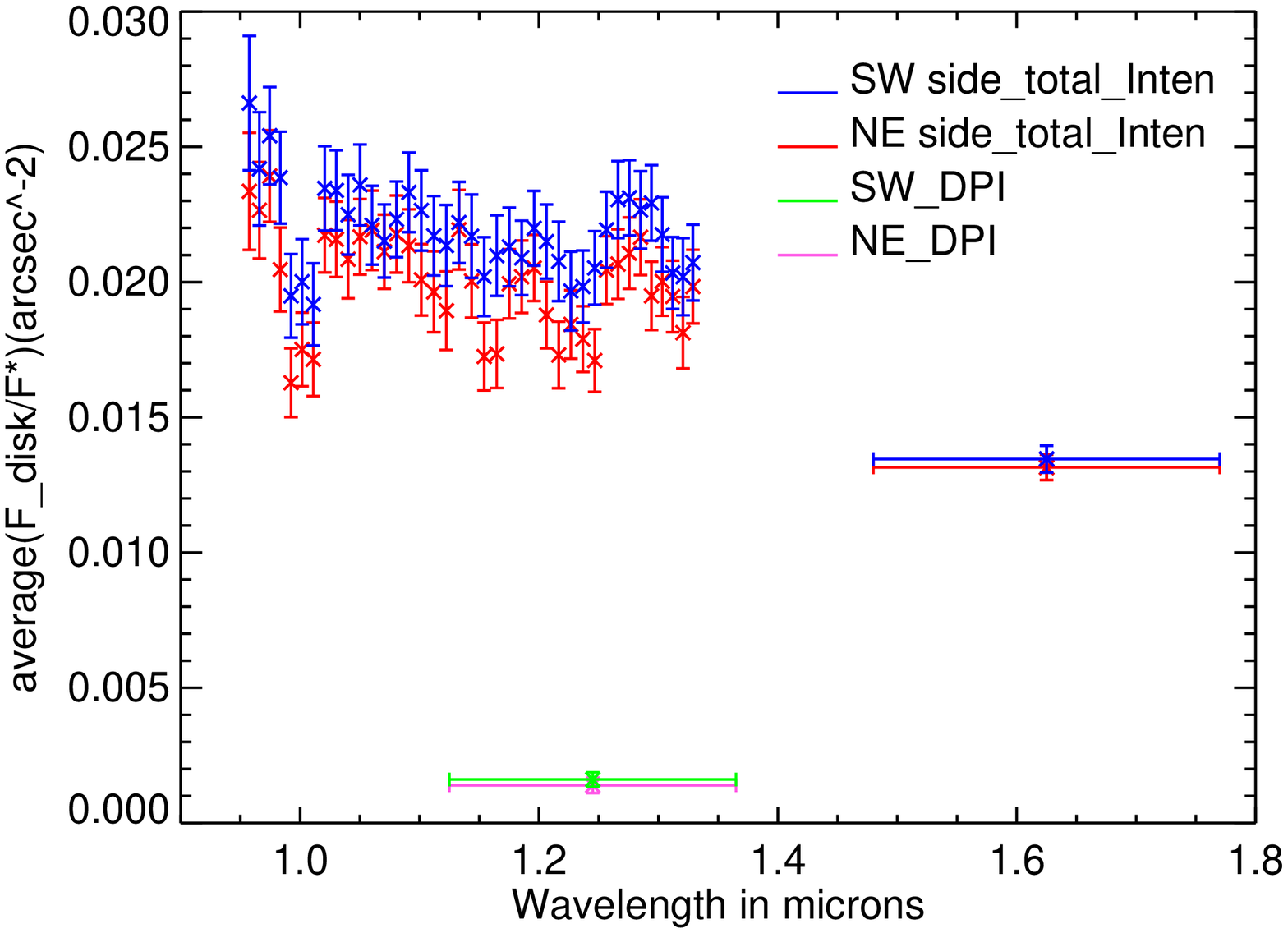}
\includegraphics[width=9cm]{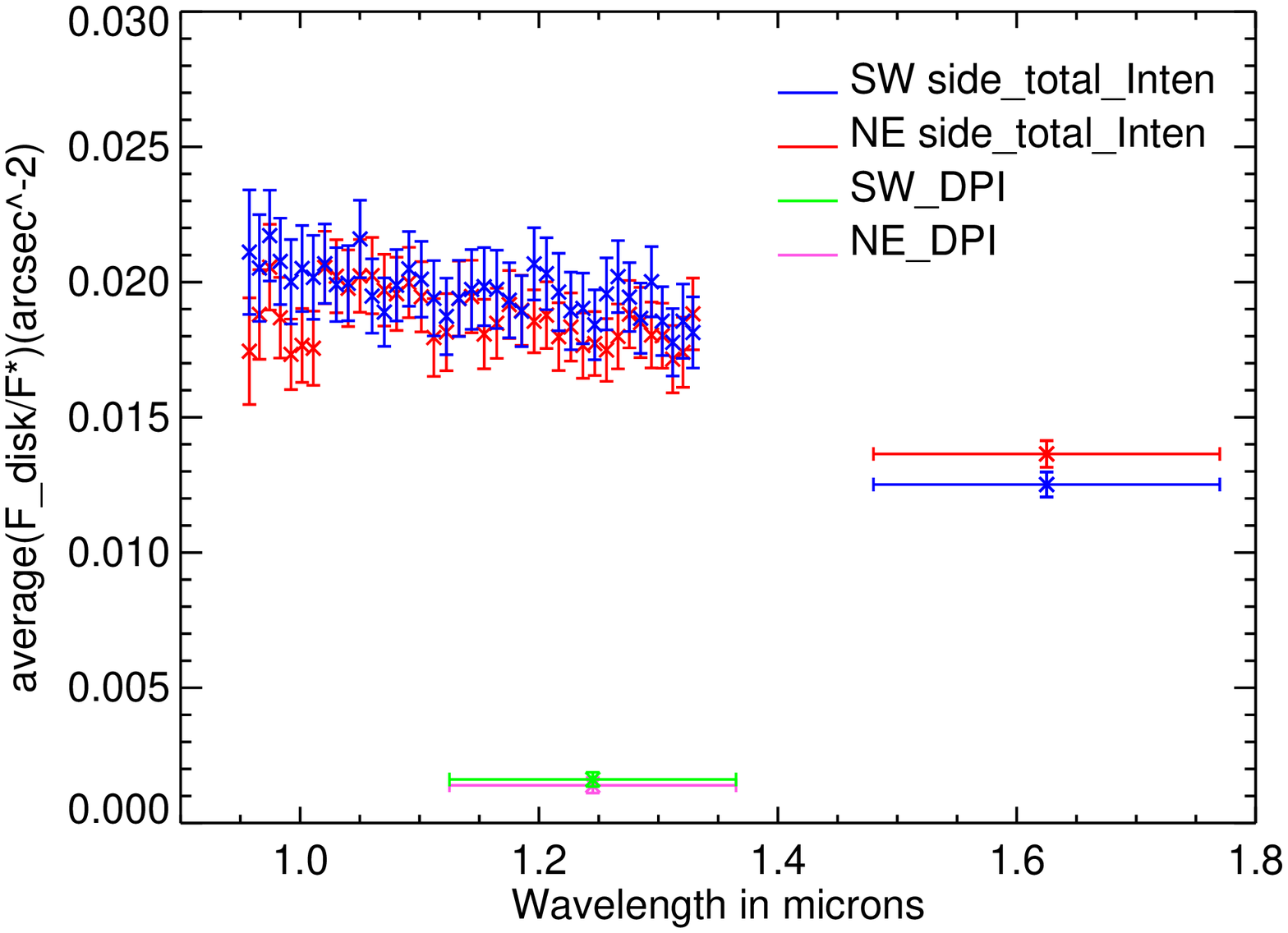}
\caption {Average spectral reflectance of HD\,32297 as measured with method 1 (left) and method 2 (right), for total intensity data in IFS YJ and IRDIS BB\_H, as well as polarimetric data in IRDIS BB\_J.}
\label{fig:spectrum}
\end{figure*}

To derive the reflectance of the grains as a function of wavelength, we converted the surface brightness profiles into contrast with respect to the star and averaged the values between  $0.2''$ and $0.8''$, 
for all spectral channels of IFS and IRDIS, as well as for polarimetric data.
Even though, the  intensity decreases by $\sim$ 1\, mag/arcsec$^2$ between 0.2 and 0.8$''$ for all spectral channels, this separation range is chosen as it corresponds to the minimal and maximal distances where the disk signal starts to dominate over the stellar halo ($>0.2''$) while encompassing the IFS field of view. 
The reflectance spectra obtained with the two methods of Sect. \ref{sec:photom_inten} are plotted in Fig. \ref{fig:spectrum}, in which the error bars are obtained from the averaged errors of the surface brightness between 0.2$''$ and 0.8$''$ for each wavelength.
The main characteristic of the reflectance, irrespective of the method used, is a slow decreasing trend with wavelength which gives the disk a gray to blue color in the YJH spectral range (Fig. \ref{fig:spectrum}). Independently of any assumptions on the grain properties, by fitting a straight line to the spectra, we measured a contrast arcsec$^{-2}$ variation of $-0.013\pm{0.002}$ per $\muup$m. 

Method 1 (Fig. \ref{fig:spectrum}, left) clearly produces more dispersion in the YJ band as a result of biases introduced by the self-subtraction estimation. Since method 2 (Fig. \ref{fig:spectrum}, right) is based on photometry extracted from models, it provides a smoother spectrum, but a slightly lower reflectance ($\sim$12 to 20\%) than method 1. For further assurance, we checked the consistency of the two methods by injecting a fake disk 90\deg to the real disk PA and performing consecutive photometry for the real disk and the fake disk with both methods. We find that the uncertainty on the  measurement between the real disk and the fake disk in Method 1 is twice that found when using Method 2. We therefore chose to use Method 2 for subsequent analyses. 

Using previous measurements from the literature to confirm the trend of the reflectance on a larger spectral range would be valuable, but photometric measurements are usually derived from various methods, and at various locations on the disk image, which makes the comparison difficult. However, in the case of HD\,32297, Ks band data were obtained by some of us using similar (but not quite identical) methods  \citep{Boccaletti2012}. In these data, the disk is only detected at stellocentric distances of 0.5-0.8$''$ for both the NE and SW sides and the surface brightness is  $\sim15\pm{0.5}$\,mag/arcsec$^2$ which translates to a contrast of $(1.09\pm{0.5})\times10^{-3}$\,arcsec$^{-2}$. 

\begin{figure*}[h]
\centering
\includegraphics[width=0.7\textwidth]{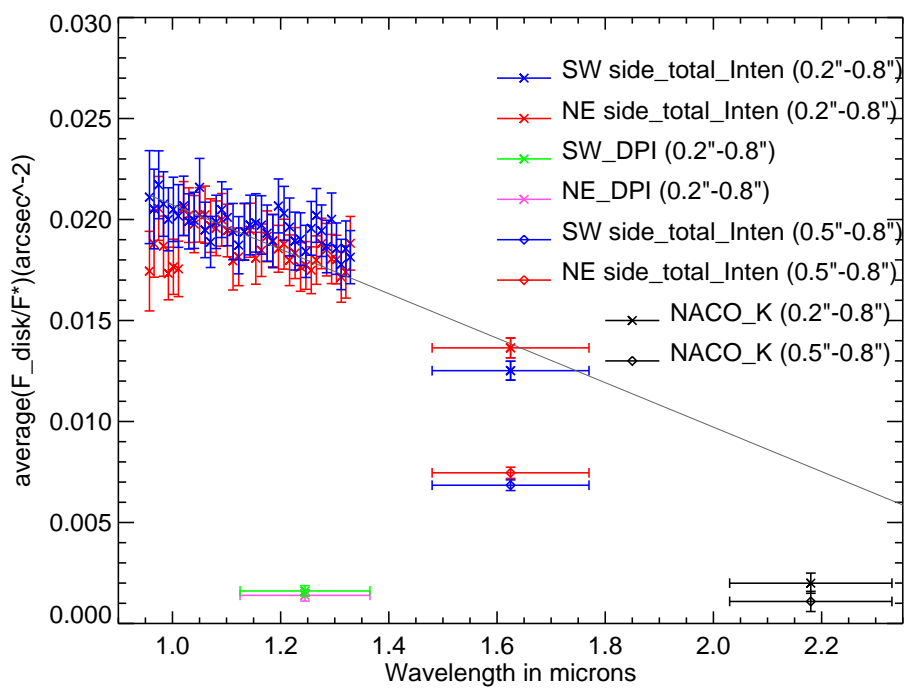}
\caption {Average spectral reflectance of HD\,32297 as measured with method 2, for total intensity data in IFS YJ (in the 0.2-0.8$''$ range) and IRDIS BB\_H (0.2-0.8$''$ and 0.5-0.8$''$), as well as polarimetric data in IRDIS BB\_J (0.2-0.8$''$). The NACO K band measurements in the range  0.5-0.8$''$  \citep{Boccaletti2012}, and an extrapolated value in the range  0.2-0.8$''$ are over-plotted. The SPHERE YJH band data are fitted with a straight line (in gray) to estimate a global slope which is $(-0.013\pm{0.002})$\,arcsec$^{-2}$ per $\muup$m.} 
\label{fig:spectrum_k}
\end{figure*}

Considering the variation of the disk intensity in SPHERE images between the two ranges of separations $0.2''-0.8''$ and $0.5''-0.8''$ at H band, we can extrapolate this Ks band contrast to $(1.99\pm{0.5})\times 10^{-3}$. 
Therefore, at first order, the photometry in the Ks band confirms the spectral slope of the reflectance (Fig.\ref{fig:spectrum_k}).{ Observing the disk at the Ks band with SPHERE would further validate the current value of the slope found}. 

The globally blue behavior of the spectrum is a key element that can help to constrain the size distribution of the grains composing the HD\,32297 disk.

\section{Grain modeling}
\label{sec:grains}

Now that we have constrained the geometry and morphology of the disk, we return to the GRaTer radiative transfer code to constrain the particle size distribution (PSD) of the dust grains.  We take as a reference morphology ($R_0\pm{dr}$) the best fit obtained for the "full disk" fit displayed in Table \ref{table:int2g} and produce synthetic spectra of the disk as a function of grain-related parameters, which we then compare to the observed spectrum in the NIR. 

We consider three different grain compositions: astro-silicates, porous astro-silicates with 80$\%$ porosity, and a mixture with 50$\%$ water ice and 50$\%$ astro-silicates. The two crucial free parameters are the minimum grain size $s_{min}$ and the index $\kappa$ of the power-law that the PSD is assumed to follow ($dn(s) \propto s^{\kappa}ds$). We explore values of $s_{min}$ between 0.1 and 10\,$\muup$m with an increment of 0.1\,$\muup$m and values of $\kappa$ ranging from -5.0 to -3.0 with an increment of 0.1.

Each model spectrum is interpolated on the forty wavelength channels (39 for IFS and 1 for IRDIS) and globally scaled to the data using the same $\chi^2$ minimization framework as in Sect. \ref{sec:model1g}.

For all wavelength channels ($i$), the reduced $\chi^2$ is measured between the average spectral reflectance $D_{i}$ and each grain model spectrum $M_{i}$ as given below

\begin{equation} 
\label{eq7}
\begin{aligned}
\chi_\nu^2 =\dfrac{1}{\nu}  \sum_{i=1}^{40}\Bigg( \dfrac{D_{i}- M_{i}(p)}{\sigma_{i}}\Bigg)^2
\end{aligned}
,\end{equation}

\noindent where $\sigma_{i}$ is given by the error bars for each data point in Fig. \ref{fig:SB} and $p$ is the parameter space. There are 40 wavelength channels ($n_{\rm data}$) and two parameters ($n_{\rm param}$; minimum grain size and distribution index), and therefore there are 38 degrees of freedom; $\nu=38$. 
The best models are selected as per 1$\sigma$ deviation of the reduced $\chi^2$ distribution, which is given by $\chi_{\nu,th}^2 = \chi_{\nu,min}^2 + \Delta{\chi_\nu^2}$, with $\Delta{\chi_\nu^2} = \sqrt{2\nu}$. 

\begin{figure*}[h]
\includegraphics[width=9cm]{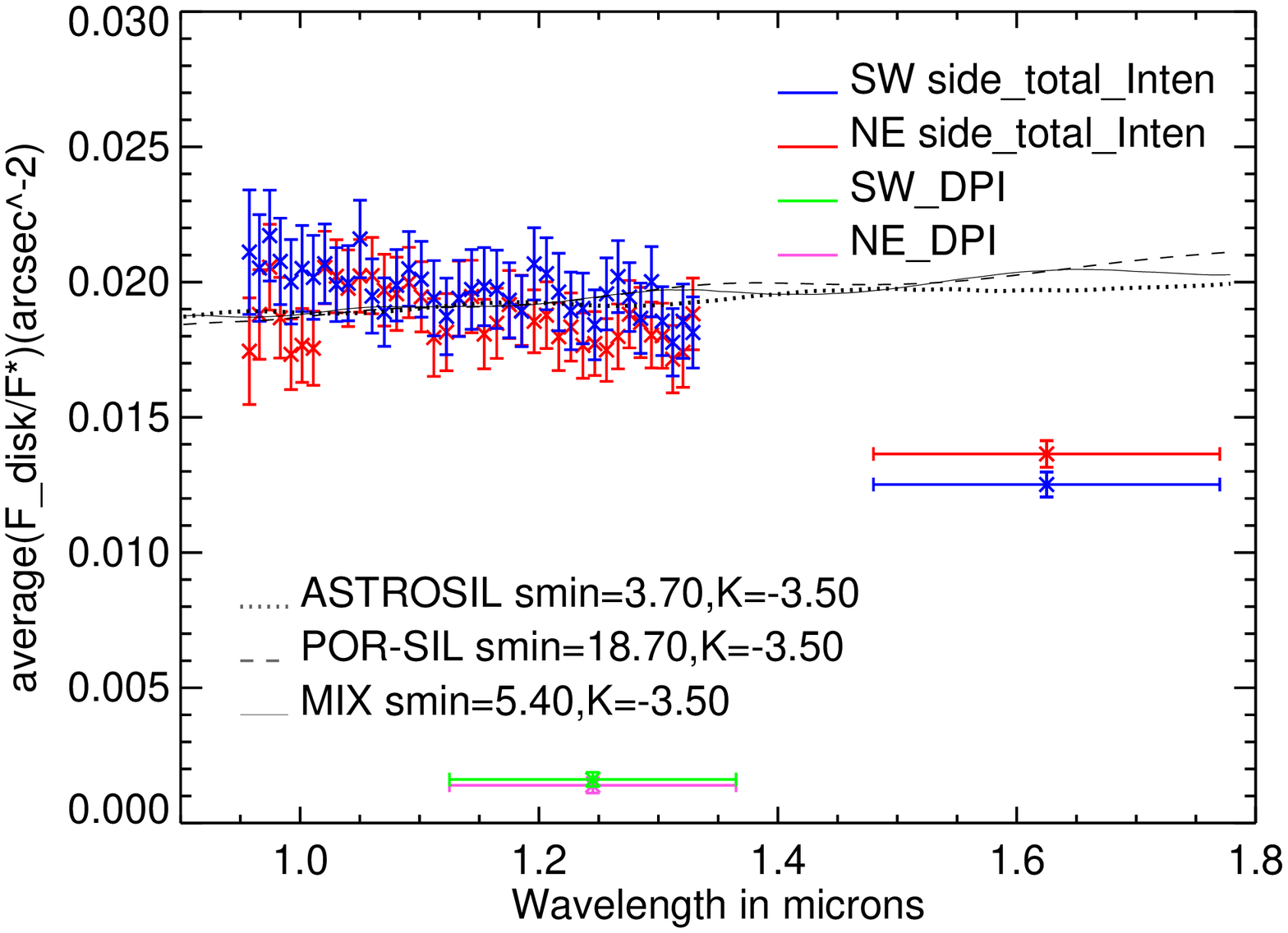}
\includegraphics[width=9cm]{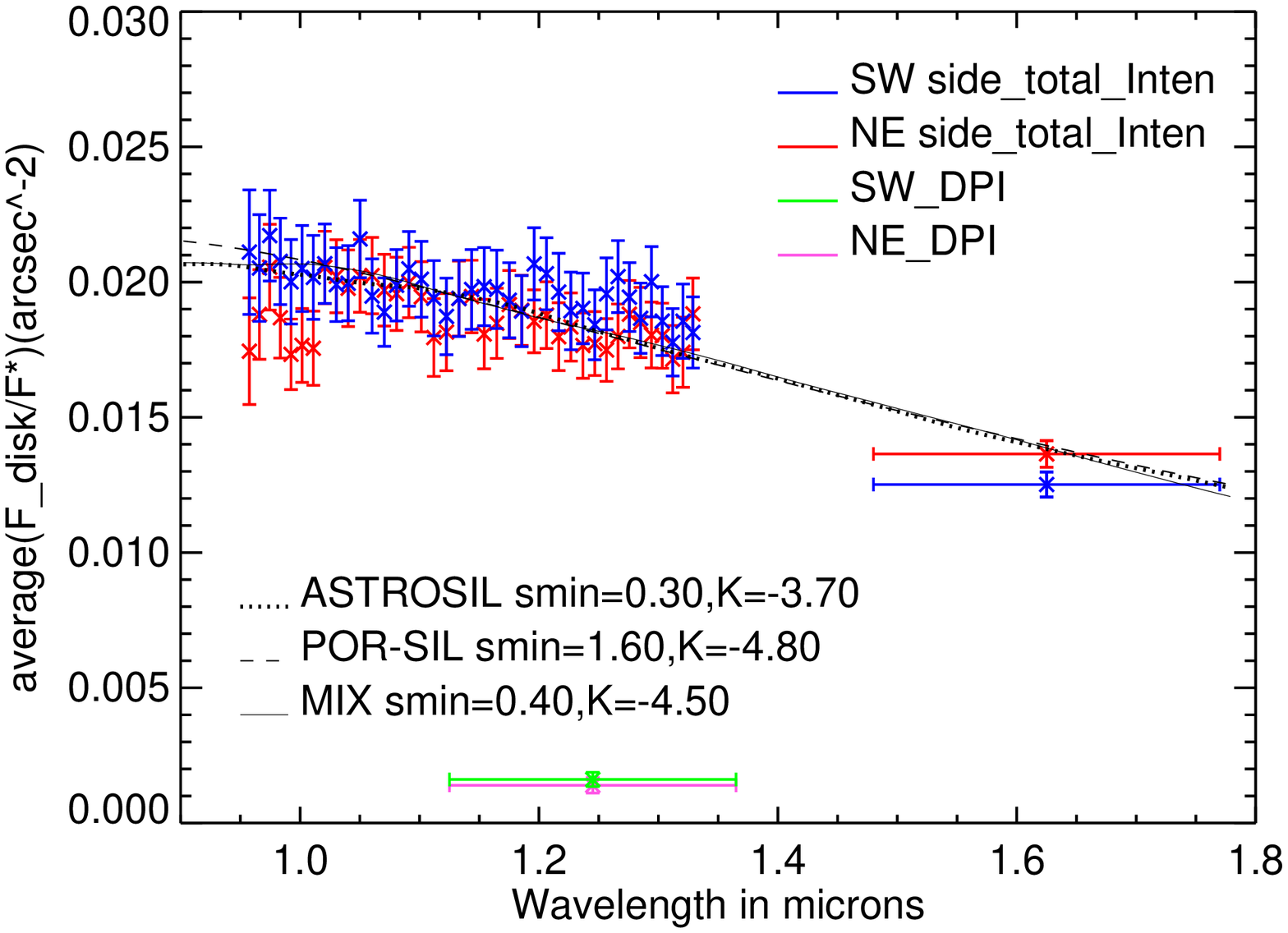}
\caption {Best fits of the reflectance spectrum (method 2) obtained for three different grain compositions, for a fiducial case where we force $s_{min}=s_{blow}$ and $\kappa= -3.5$ (left), and when $s_{min}$ and $\kappa$ are free parameters (right).
}
\label{fig:fitSpectrum}
\end{figure*}

The reflectance spectrum corresponding to the best fits obtained for the three considered compositions is displayed in Fig. \ref{fig:fitSpectrum} (right) and the best-fit parameters are provided in Table \ref{table:grain}. An important result is that, for all considered compositions, $s_{min}$ is well below the blow-out limit size $s_{blow}$. We indeed always have $s_{min}/s_{blow} \leq 0.085$, where we derive $s_{blow}$ in gm/cm$^3$ using the prescription by \cite{Wyatt2008}:

\begin{equation}
\label{eq7a}
\begin{aligned}
    s_{blow}=0.8\dfrac{{\rm L}_*}{{\rm M}_*}\dfrac{2.7}{\rho}
    \end{aligned}
,\end{equation}

\noindent where L$_*$ and  M$_*$ are the stellar luminosity and mass expressed in solar values, and $\rho$ is the bulk density of the material (given in Table~\ref{table:grain}). We take L$_*= (8.4\pm{0.2})$  $L_{\odot}$ \citep{Moor2017} and M$_* =1.8$M$_\odot$ \citep{Kalas2005}.   

\begin{table*}[h]
\centering
\caption{Parameters of the grains and their size distribution that generate the best fit to the spectrum} 
\begin{tabular}{c | c|c |c | c | c | c |c  }     % 7 columns  
\hline \hline  

  Grain type & volume ratio &density g/cm$^3$& $\kappa$ & $s_{min}$ & $s_{blow}$ & $s_{min}/s_{blow}$ & $\chi^2$\\
\hline
  Astro-silicate          & -   & 2.7  & $-3.79\pm{0.34}$ & $0.30\pm{0.02}$ & 3.7 & 0.081 & $0.471\pm{0.084}$ \\ 
  Vacuum+astro-silicate  & 4:1 & 0.54 & $-4.24\pm{0.60}$ & $1.59\pm{0.11}$ & 18.7 & 0.085 & $0.628\pm{0.114}$ \\
  Astro-silicate+water ice & 1:1 & 1.85 & $-4.65\pm{0.35}$ & $0.44\pm{0.03}$ & 5.4 & 0.081 & $0.502\pm{0.091}$ \\
\hline \hline                 

\end{tabular}
\label{table:grain} 

\end{table*}

As for the best fit of the slope (Fig.\ref{fig:fitSpectrum}, right) of the size distribution, we find $\kappa=-3.79\pm{0.34}$ for astro-silicates, which is relatively close to the slope expected for collisional steady states \citep[between -3.6 and -3.7; see][and references therein]{Gaspar2012}. For the other two compositions, we find slightly steeper PSDs, with $\kappa\sim-4.5$. For all three slopes, it is important to stress that the geometrical cross section, and thus the flux, is dominated by the smallest grains in the PSDs, that is, those close to $s_{min}$ \citep{Thebault2019}. Also, $s_{min}=2.2\muup$m, the minimum size found by \cite{Donaldson2013}, also corresponds to $s<s_{blow}$ grains given the very high 90\% porosity they assumed. To further stress the crucial role of unbound $s\leq s_{blow}$ grains, we display in Fig.\ref{fig:fitSpectrum} (left) the best fits that would be obtained when considering a PSD stopping at $s_{min}=s_{blow}$ and a canonical slope with $\kappa = -3.5$. As can be clearly seen, in the absence of the unbound grains no satisfying match can be obtained, in particular regarding the blue slope of the reflectance spectrum. 

These results, especially those regarding the minimum grain size, only weakly depend on the morphology assumed for the disk. Taking a size for the planetesimal belt other than that of Table \ref{table:int2g} indeed leads to relatively similar results. This is expected, as the slope of the reflectance spectrum is essentially imposed by the size-dependence of the scattering coefficient $Q_{sca}$, which does not depend on the location of the grains with respect to the star. The scattering anisotropy parameters $g$ do depend on this location, but under the assumption made in Sect.\ref{sec:model} that there is no size dependence for $g$, this would not translate into a different slope of the reflectance spectra for a given PSD. Therefore, we note that at this stage it is safe to derive synthetic spectra without a dependence of anisotropic scattering parameter.

\section{Discussion}
\label{sec:discuss}

\subsection{Comparison of geometrical parameters to millimeter observations by ALMA}

From the analysis presented in Sect. \ref{sec:model_pol} we find that the disk is best described as a relatively narrow belt peaking at $132.3\pm{6.2}$ au (according to polarimetric data), which agrees with previous scattered light observations \citep[after the correction of the distance from the new Gaia measurements]{Currie2012,Boccaletti2012}. HD\,32297 was also observed recently with ALMA in the dust continuum at 1.3\,mm by MG18. However, the beam size was about ${0.76''\times0.51''}$ which is more than ten times larger than the SPHERE angular resolution, meaning that the {ALMA} resolution is significantly poorer. 

While considering a geometrical model that is similar to GRaTer, MG18 concluded that the disk is rather broad, extending from $78\pm8$\,au to $122\pm3$\,au, with a surface density rising as $r^2$. The peak in density occurs slightly closer-in than in the SPHERE images, but this discrepancy could be the result of the angular resolution. In any case, the radial dependency of the surface density, even if relatively poorly constrained in our case, is considerably steeper with SPHERE.    

 Beyond the planetesimal ring at 122\,au, MG18 observed a halo extending out to $440\pm{32}$\,au, where the surface density decreases as $r^{-6}$. This is consistent with the SPHERE image in which we observe dust scattering as far as 3.3$''$ (equivalent to 440\,au), with a comparable radial decrease in density. MG18 do not report any asymmetry in this outer part while the aforementioned concavity is  obvious
 at shorter wavelengths. This could again be a resolution effect
 or because of the sensitivity of ALMA to bigger grains. 
 Nevertheless, subtracting an axi-symmetrical disk model from the ALMA image leaves residuals co-located with the region where the concavity is detected by SPHERE and HST. The presence of millimeter grains at such stellocentric distances {would, however, be at odds with the} expected mechanism creating such concavity.
 For the same reason, our value for the disk inclination is significantly more precise ($88.2\pm{0.3}$\deg) than the one derived by MG18 ($83.6\degb$~$^{+4.6}_{-0.4}$). 
 
In order to check the compatibility of the ALMA and SPHERE models, we used the MG18 parameters $\alpha_{in}$, $\alpha_{out}$, $R_{in}$, $R_{out}$, and took the other parameters  ($g_1,g_2,w_1,h$) from our best-fit values (Table.~\ref{table:comp_mm}). 
{As a result, we obtained a much larger $\chi^{2}$ value of 8.5 for total intensity compared to our best fit. For polarimetry, $\chi^2=3.6,$ which is not far from the value we find for our best-fit model. This is because $R_0=122$ au, as used in MG18, is close to that of our polarimetric best-fit model $R_0=125$ au.} We also explored models with $g_1,g_2,w_1$ as free parameters corresponding to the parameter space of Sect \ref{sec:model2g} for total intensity and Sect \ref{sec:model_pol} for polarimetry in order to restrict possible degeneracies between dust density distribution corresponding to the ALMA values and phase function. {The attempt resulted in larger $\chi^{2}$ (8.3 for total intensity and 3.6 for polarimetry) values compared to our best fit indicating that the ALMA model does not match the SPHERE image. The disagreement between the models derived from these two instruments could stem from one of two factors, or both. The first is the angular resolution as mentioned above. The second is that ALMA and SPHERE probe different grain size of which the dynamics can be governed by different processes resulting in two distinct spatial distributions.}

\begin{table}[h]   
\caption{Parameters used to create models comparable to millimeter observations by ALMA. Here, 
$R_{in}$ and $R_{out}$ are the inner and outer edges of the disk
}

\begin{tabular}{c| c c  }     % 7 columns  
\hline \hline  
  Parameters &Total intensity&Polarimetry \\
\hline
  Inclination i (\deg)  & 88.0 & 88.5 \\
  $R_0$ & 122 & 122 \\
  $\alpha_{in}$ & 2& 10 \\
  $\alpha_{out}$ & -6 & -4 \\
  $R_{in}$ &78 & 78 \\
  $R_{out}$ & 440 & 440 \\
  $g_1$ & 0.7 & 0.8 \\
  $g_2$ & -0.4 & 0.0 \\
  $w_1$ & 0.80 & 1.00 \\
  $h=H_{0}/R$ & 0.020 & 0.020 \\
  $\chi^{2}$ & {8.53} & {3.64} \\
\hline \hline                 
\end{tabular}
\label{table:comp_mm} 
\end{table}

\subsection{Disk color and sub-micron grains}
\subsubsection{Quantity of sub-micron grains produced naturally in debris discs}

Our new observations and analysis confirm two striking characteristics of the HD\,32297 disk: the blue color of the spectrum in the NIR and the significant presence of tiny grains much smaller than $s_{blow}$. Moreover, we confirm that there is an intrinsic coupling between these two characteristics, as was also inferred for the HD\,15115 \citep{Debes2008} or AU\,Mic systems \citep{Augereau2006, Fitzgerald2007b}.

This link between a blue NIR spectrum and sub-micron grains has been quantitatively investigated in the recent study by \cite{Thebault2019}. This numerical exploration shows that for bright debris disks with high fractional luminosity $f_d$, a  collisional cascade at steady-state can "naturally" produce a level of unbound sub-micron grains that is high enough to lead to a blue slope of the spectrum in the NIR. This is because for bright and dense disks, the drop in grain number density at the $s= s_{blow}$ frontier, which is to a first order $\propto 1/f_d$, is much less pronounced than for fainter systems. For a very bright disk with $f_d=5\times 10^{-3}$ comparable to that of HD\,32297, \cite{Thebault2019} found a profile of the relative NIR $L_d/L*$ spectrum that is qualitatively similar to the one obtained here (see Fig. 13 of that paper). However, the blue slope they obtained is not as steep as in the present case, with a flux ratio between the $\lambda=1\muup$m and $\lambda=1.6\muup$m fluxes that is $\sim 1.1$, as compared to $\sim 1.6$ here. 

This could indicate an additional source of sub-micron grains, which cannot be explained by the steady-state collisional evolution of the system. One possible cause could be the so-called collisional "avalanche" mechanism \citep{Grigorieva2007b,Thebault2018}, initiated by the break-up of a large planetesimal closer to the star, which releases large amounts of unbound dust grains that then trigger a collisional chain-reaction as they sandblast at very high velocity through a dense outer disk. The ideal case for an avalanche-producing system is a double-belt configuration, with an inner belt (where the large planetesimal breaks up) at $\sim$1-10\,au with $f_d\gtrsim 10^{-4}$, and a bright outer belt with $f_d\gtrsim 10^{-3}$ \citep{Thebault2018}.This could match the structure of HD\,32297, for which an inner belt of brightness $f_d\sim6\times 10^{-4}$ has been inferred by \cite{Donaldson2013}, even though the reality of this inner belt is still debated \citep[e.g.,][]{Kennedy14}. {Observational confirmation of an inner belt would need to achieve contrasts significantly higher (a factor of 10) than those currently feasible with SPHERE. } 
In this case, the level of $s\leq s_{blow}$ grains would vary stochastically, on a timescale $t_{av}$ that is roughly a third of the typical dynamical timescale in the disk \citep{Thebault2018}. This would, however, correspond here to $t_{av}\sim 300-400\,$yrs, much too long to be observationally monitored. Moreover, it is not guaranteed that the rate at which large planetesimals break up in the inner regions is high enough for such an event to be likely to be witnessed \citep[see discussion in][]{Thebault2018}.

\subsubsection{Effect due to gas on the presence of sub-micron grains}

Another possibility is that the system is able to retain $s\leq s_{blow}$ grains significantly longer than the radiation pressure blow-out time. If there is enough gas in the system, gas drag could act to significantly increase the time for an unbound grain to leave the system. To check that, we first compute the stopping time for the case where grains are bound to their host star, equal to \citep{Takeuchi2001}

\begin{equation}
\label{eqst}
\begin{aligned}
    T_{\rm sb} \sim 2 \left( \frac{\rho}{1.5\,{{\rm g/cm}^3}} \right) \left( \frac{s}{1\, \mu{\rm m}} \right) \left( \frac{M_{\rm gas}}{0.1\, {\rm M}_\oplus} \right)^{-1},
    \end{aligned}
\end{equation}

\noindent where we assume M$_\star=1.8$ M$_\odot$, $R=130$ au, $\Delta R=50$ au to be consistent with results from Sect.~\ref{sec:model}. We also fix the gas temperature to 30K and its mean molecular weight to 28 based on \citet{Cataldi2019}, where they show that the gas mass is dominated by CO rather than carbon in HD\,32297. Accounting for the observed neutral and ionized carbon \citep[in addition to CO][submitted]{Moor2019}, the total gas mass barely goes above 0.1 M$_\oplus$. If accounting for potential CO$_2$ or water being released from planetesimals at the same time as CO (and thus producing extra oxygen not coming from CO and some extra hydrogen), and assuming a solar-system comet-like composition \citep[e.g.,][]{Kral2016}, the total gas mass could go up to 0.5 M$_\oplus$\footnote{We note that if the gas were primordial, the total gas mass (accounting for extra H$_2$) could go up to $10^2$ M$_\oplus$, but this is probably not the case as shown in \citet{Kral2017,Kral2018}}.

Therefore, small bound grains close to the blow-out limit (i.e., 1-10 $\mu$m, see Table \ref{table:grain}) will have a stopping time close to 1 and will be affected by gas drag over a few orbital periods before they have time to collisionally deplete. The smallest bound grains will likely (depending on the gas-pressure gradient) move outwards before being collisionally destroyed \citep{Takeuchi2001} and will therefore be present for longer than usually assumed in a standard size distribution \citep[e.g.,][]{Kral2013}.

For unbound grains, Eq.~\ref{eqst} needs to be adjusted. The velocity of unbound grains can reach $\sqrt{2(\beta-1)}$ of the Keplerian velocity at which they are released initially, which would increase the velocity difference between gas and dust, hence entering the Stokes regime of drag \citep[rather than the Epstein regime,][]{Takeuchi2001}. On top of that, the stopping time of Eq.~\ref{eqst} is calculated over an orbital period and unbound grains travel almost radially over $\Delta R$, the width of the disk. Therefore, we calculate a new dimensionless Stokes stopping time for unbound grains $T_{\rm su}$ \citep[based on][]{Takeuchi2001}, scaled by the crossing time over $\Delta R$ and therefore equal to $T_{\rm sb} [2 c_s/v_K] [R/\Delta R$], where $v_K$ is the Keplerian velocity and $c_s$ the sound speed:

\begin{equation}
\label{eqst2}
\begin{aligned}
    T_{\rm su} \sim 0.03 \left( \frac{\rho}{1.5\,{{\rm g/cm}^3}} \right) \left( \frac{s}{0.1\, \mu{\rm m}} \right) \left( \frac{M_{\rm gas}}{0.1\, {\rm M}_\oplus} \right)^{-1},
    \end{aligned}
\end{equation}

\noindent meaning that 0.1 $\mu$m grains will be slowed down significantly before they have time to leave the disk\footnote{If grains become slow enough because of gas drag and come back to the Epstein regime, we calculate that for $\beta$ values between 1 and 5 \citep[realistic for an A6V star, see][]{Thebault2019}, the unbound stopping times (over the crossing time) are roughly one to two orders of magnitude larger than derived in Eq.~\ref{eqst}, meaning that even in the Epstein regime, grains of 0.1 $\mu$m will also have stopping times close to 1-10, i.e., gas will have time to substantially brake unbound grains before they leave the disk.}. The exact orbit of unbound grains interacting with gas is time dependent and is not derived here as it goes beyond the scope of this paper, but generally speaking, an unbound grain would start on a very hyperbolic orbit and eventually be circularized around the star. These grains will then accumulate before being destroyed collisionally. Therefore, we find tantalizing evidence that gas observed in this system may be able to explain the blue color of the disk by allowing small unbound grains to be present for longer. 
 
 \subsection{Point-source detection limit}
\begin{figure}[h]
\includegraphics[clip, trim={0 0 0 0},width=9cm]{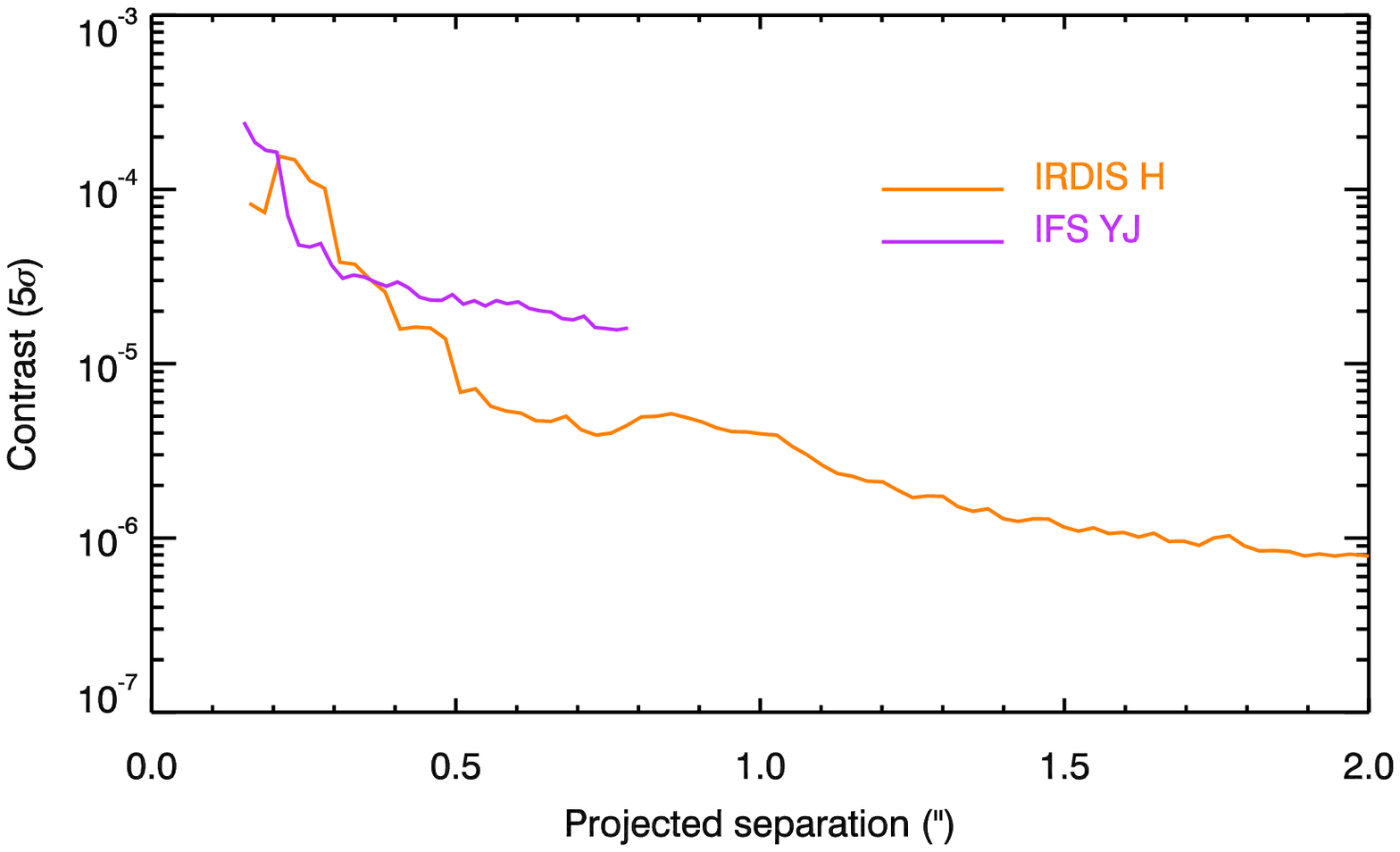}
\\
\includegraphics[width=9cm]{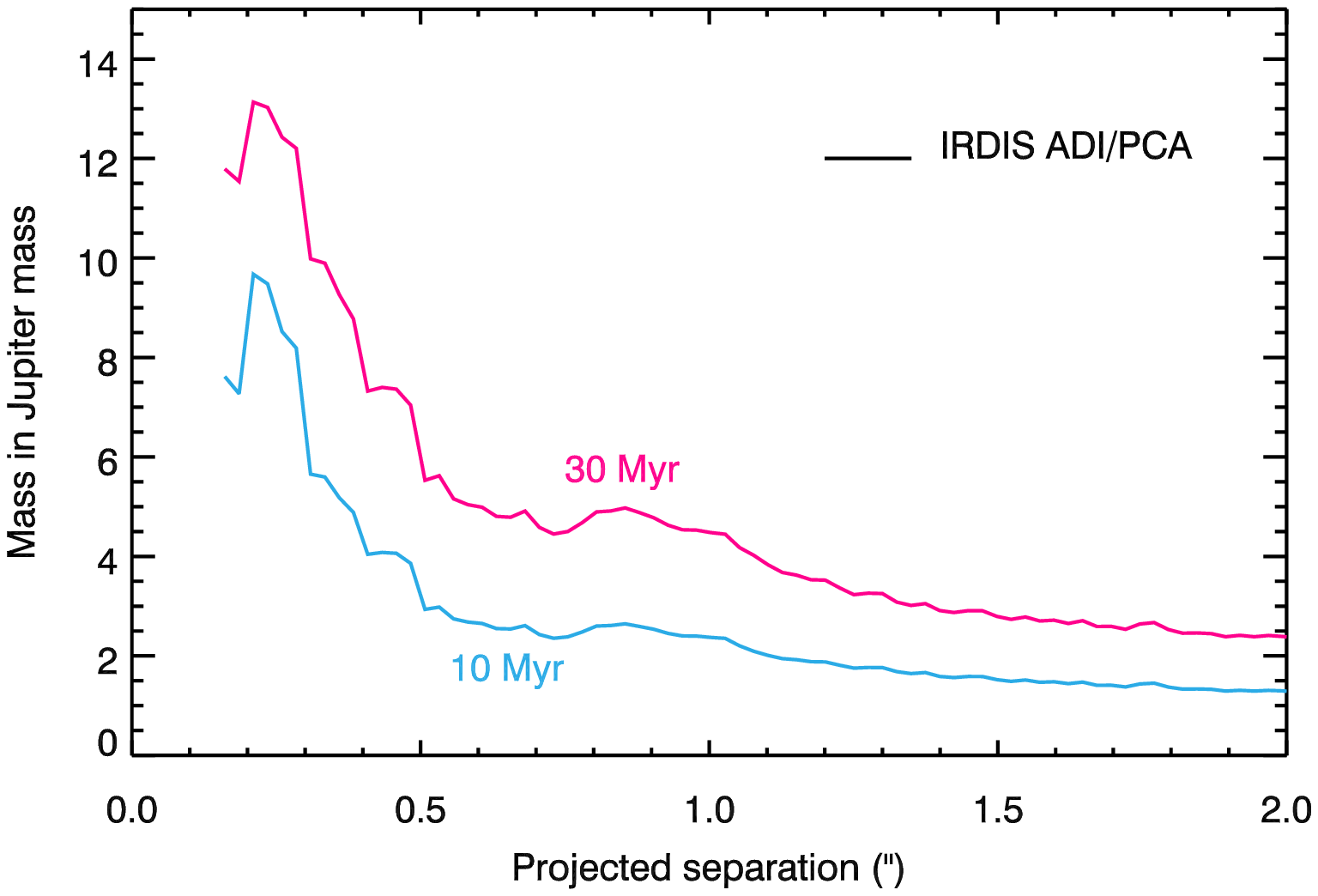}
\caption {
Limits of detection in contrast (top) for IRDIS (solid line) and IFS (dashed line), and converted into Jovian masses (bottom, IRDIS only) for two age assumptions (10 and 30 Myr), using the COND evolutionary model.}
\label{fig:contrast_curve}
\end{figure}

We do not detect any point source in the entire IRDIS field of view. The contrast curves at 5$\sigma$ are measured with the SpeCal pipeline \citep{Galicher2018} for KLIP-reduced data (Fig. \ref{fig:contrast_curve}). 
IRDIS provides contrasts of about $10^{-5}$ at 0.5$''$ and $10^{-6}$ at 1$''$. The IFS contrasts are similar or slightly better at $<0.4''$ and then degrade for larger separations. For this reason, we considered the IRDIS contrast curve only to derive the limit of detection in terms of mass. Figure \ref{fig:contrast_curve} displays the limit of detection using the COND model \citep{Allard2001}, and assuming two possible ages of the system: 10 Myr and 30 Myr.
We note that if the system were 10 Myr old we should be able to detect a planet of 3 Jupiter masses at a separation of 0.5$''$ (and respectively {1.3}$''$ for 30 Myr). The values below 1\,M$_\mathrm{J}$ are unreliable in the framework of the COND model. 
{Since the noise is estimated azimuthally in SpeCal, the disk itself contributes 40\% to the contrast curve in the range $0.2''-1.2''$}.

\section{Conclusions}
\label{conclusion}

The findings of this paper can be summarized as follows:

   \begin{itemize}
       \item 
       We observed the debris disk of HD\,32297 in the NIR in the Y, J, and H bands out to stellocentric distances of 3.3$''$, and for the first time as close as 0.15$''$. We obtained both total intensity and polarimetric images as well. 
       
       \item 
        At large separations, the disk is characterized by a concavity as reported by \cite{Schneider2014}. At shorter separations ($<1''$), a bow-like shape is reminiscent of a very inclined belt of which we see mostly one side (northwest) due to the forward scattering by the grains. Noticeably, we were able to detect the back side of the disk which we modeled using two HG phase functions. 
       This feature is not observed in polarimetric data possibly due to low S/N.
       
       \item Upon first inspection, the disk appears to be symmetrical in NE and SW sides and has no gapped structure in contrast to the claims of \cite{Asensio-Torres2016}. 
       This is confirmed unambiguously with our photometric study in which the surface brightness profiles do not show any significant brightness asymmetry between the two sides, or any gap.
       
       \item We present two methods for extracting the photometry of inclined disks observed in total intensity. The first method includes the estimation of the ADI self-subtraction in model images and accounting for this bias into the data. {We find that this method can induce some irregularities depending on the accuracy of measurement of the ADI self-subtraction.} In the second method we calculate a scaling factor between the ADI processed data and its corresponding model. The scaled model is used to measure the photometry instead of the data. As a drawback any departure from the model is not represented in the measurements.
       
       \item Comparing total intensity and polarimetry in the J band we derived a polarisation fraction of about $15\%$ which is in accordance with other debris disks.
       
       \item From photometric measurements obtained in 40 spectral channels we obtained an average spectral reflectance and conclude that the disk is {`gray to blue' color }in the YJH spectral range. 
     Using a radiative transfer module in GRaTer we were able to compare this measured reflectance with those expected for a variety of grain sizes and compositions. We found that irrespective of the composition, grains should be significantly smaller than the corresponding blowout size (sub-micron size for astrosilicates).  
       
       \item Finally, we discussed that the presence of the small grains and the associated blue color of the disk can originate from a combination of several physical processes, including steady-state collisional evolution and the avalanche process. 
       Given the amount of gas in this system \citep{Greaves2016,Cataldi2019}, we also found that 
      the gas drag can retain smaller unbound grains over a longer timescale.
       
   \end{itemize}
   
  HD\,32297 is amongst the very few known bright and extended debris disks with gas. The SPHERE observations are of unprecedented quality allowing the detection of this disk at high S/N in all spectral channels, and strong constraints to be derived on the grain properties. 
  Confirming the trend of the spectral reflectance would require additional SPHERE observations in the K band in total intensity as well as  polarimetric data in the H and K bands. 
  It would be interesting to perform the very same type of observations and data analysis for other gas-rich debris disks and investigate if they share similarities with HD\,32297 as an attempt to understand whether the presence of gas can fully explain the dust size distribution.  

\begin{acknowledgements}
French co-authors acknowledge financial support from the Programme National de Plan{\'e}tologie (PNP) of CNRS-INSU in France. The project is supported by CNRS, by the Agence Nationale de la Recherche (ANR-14-CE33-0018).
Finally, this work has made use of the SPHERE Data Centre, jointly operated by OSUG/IPAG (Grenoble), PYTHEAS/LAM/CESAM (Marseille), OCA/Lagrange (Nice) and Observatoire de Paris/LESIA (Paris). J.Ma. acknowledges support for this work was provided by NASA through the NASA Hubble Fellowship grant HST-HF2-51414.001 awarded by the Space Telescope Science Institute, which is operated by the Association of Universities for Research in Astronomy, Inc., for NASA, under contract NAS5-26555.
      
\end{acknowledgements}

%-------------------------------------------------------------------

\bibliographystyle{aa} % style aa.bst
\bibliography{all} % your references Yourfile.bib
\end{document}